\input harvmac
\noblackbox 
\ifx\answ\bigans
\magnification=1200\baselineskip=14pt plus 2pt minus 1pt
\else\baselineskip=16pt 
\fi


\def\INT{{\int\limits_0^1\hskip-0.1cm dx\hskip-0.15cm\int\limits_0^1
\hskip-0.1cm dy\hskip-0.15cm\int\limits_0^1 \hskip-0.1cm dz}}

\hyphenation{Gauss--ian}

\def\gym{g_{Y\! M}}

\def\ap{\alpha'}

\def\eps{\epsilon}
\def\al{\alpha}

\def\si{\sigma}

\def\bet{\beta}

\def\Si{{\Sigma}}

\def\comment#1{{}}

\def\slashchar#1{\setbox0=\hbox{$#1$}           
   \dimen0=\wd0                                 
   \setbox1=\hbox{/} \dimen1=\wd1               
   \ifdim\dimen0>\dimen1                        
      \rlap{\hbox to \dimen0{\hfil/\hfil}}      
      #1                                        
   \else                                        
      \rlap{\hbox to \dimen1{\hfil$#1$\hfil}}   
      /                                         
   \fi}
\newif\ifnref
\def\rrr#1#2{\relax\ifnref\nref#1{#2}\else\ref#1{#2}\fi}
\def\ldf#1#2{\begingroup\obeylines
\gdef#1{\rrr{#1}{#2}}\endgroup\unskip}

\def\doubref#1#2{\refs{{#1},{#2} }}
\def\threeref#1#2#3{\refs{{#1},{#2},{#3} }}

\nreffalse

\def\lref{\ldf}

\input epsf
\def\figin{\epsfcheck\figin}\def\figins{\epsfcheck\figins}
\def\epsfcheck{\ifx\epsfbox\UnDeFiNeD
\message{(NO epsf.tex, FIGURES WILL BE IGNORED)}
\gdef\figin##1{\vskip2in}\gdef\figins##1{\hskip.5in}
\else\message{(FIGURES WILL BE INCLUDED)}%
\gdef\figin##1{##1}\gdef\figins##1{##1}\fi}
\def\DefWarn#1{}
\def\figinsert{\goodbreak\midinsert}
\def\ifig#1#2#3{\DefWarn#1\xdef#1{fig.~\the\figno}
\writedef{#1\leftbracket fig.\noexpand~\the\figno}%
\figinsert\figin{\centerline{#3}}\medskip\centerline{\vbox{\baselineskip12pt
\advance\hsize by -1truein\noindent\footnotefont{\bf Fig.~\the\figno } #2}}
\bigskip\endinsert\global\advance\figno by1}


\def\appA{A}
\def\appB{B}

\def\tilde{\widetilde}

\def\h {{1\over 2}}

\def\ov {\overline}
\def\o {\over}
\def\fc#1#2{{#1 \o #2}}

\def\IR{ {\bf R}}


\def\br{\hfill\break}

\def\lf {\left}
\def\ri {\right}

\def\p {\partial}

  \def\Nc{{\cal N}}

\def\Ic {{\cal I}}


\lref\LHC{D. L\"ust, S.~Stieberger and T.R.~Taylor,
``The LHC String Hunter Companion'', MPP--2008--05, to appear.}

\lref\BI{
S.~Stieberger and T.R.~Taylor,
``Non-Abelian Born-Infeld action and type I - heterotic duality.  I:
Heterotic $F^6$ terms at two loops,''
  Nucl.\ Phys.\  B {\bf 647}, 49 (2002)
  [arXiv:hep-th/0207026];
``Non-Abelian Born-Infeld action and type I - heterotic duality. II:
Nonrenormalization theorems,''
  Nucl.\ Phys.\  B {\bf 648}, 3 (2003)
  [arXiv:hep-th/0209064].
}
\lref\August{S.~Stieberger and T.R.~Taylor,
``Supersymmetry Relations and MHV Amplitudes in Superstring Theory,''
  Nucl.\ Phys.\  B {\bf 793}, 83 (2008)
  [arXiv:0708.0574 [hep-th]].
}
\lref\BrittoAP{
  R.~Britto, F.~Cachazo and B.~Feng,
  ``New recursion relations for tree amplitudes of gluons,''
  Nucl.\ Phys.\  B {\bf 715}, 499 (2005)
  [arXiv:hep-th/0412308].
}
\lref\simmons{E.H.~Simmons,
  ``Dimension-Six gluon operators as probes of new physics,''
  Phys.\ Lett.\  B {\bf 226}, 132 (1989); ``Higher dimension gluon operators and hadronic scattering,''
  Phys.\ Lett.\  B {\bf 246}, 471 (1990);\br
    P.L.~Cho and E.~H.~Simmons,
  ``Searching for G3 in t anti-t production,''
  Phys.\ Rev.\  D {\bf 51}, 2360 (1995);\br
    H.K.~Dreiner, A.~Duff and D.~Zeppenfeld,
  ``How well do we know the three gluon vertex?,''
  Phys.\ Lett.\  B {\bf 282}, 441 (1992);\br
  L.J.~Dixon and Y.~Shadmi,
  ``Testing gluon selfinteractions in three jet events at hadron colliders,''
  Nucl.\ Phys.\  B {\bf 423}, 3 (1994)
  [Erratum-ibid.\  B {\bf 452}, 724 (1995)]
  [arXiv:hep-ph/9312363].}
\lref\BrittoFQ{
  R.~Britto, F.~Cachazo, B.~Feng and E.~Witten,
  ``Direct proof of tree-level recursion relation in Yang-Mills theory,''
  Phys.\ Rev.\ Lett.\  {\bf 94}, 181602 (2005)
  [arXiv:hep-th/0501052].
}
\lref\CFT{
  D.~Friedan, E.J.~Martinec and S.H.~Shenker,
``Conformal Invariance, Supersymmetry And String Theory,''
  Nucl.\ Phys.\  B {\bf 271}, 93 (1986);\br
J.~Cohn, D.~Friedan, Z.a.~Qiu and S.H.~Shenker,
``Covariant Quantization Of Supersymmetric String Theories: The Spinor
  Field Of The Ramond-Neveu-Schwarz Model,''
  Nucl.\ Phys.\  B {\bf 278}, 577 (1986).
}

\lref\Grisaru{
  M.T.~Grisaru, H.N.~Pendleton and P.~van Nieuwenhuizen,
``Supergravity And The S~Matrix,''
  Phys.\ Rev.\  D {\bf 15}, 996 (1977);\br
M.T.~Grisaru and H.N.~Pendleton,
``Some Properties Of Scattering Amplitudes In Supersymmetric Theories,''
  Nucl.\ Phys.\  B {\bf 124}, 81 (1977).
}

\lref\KLLSW{
  V.A.~Kostelecky, O.~Lechtenfeld, W.~Lerche, S.~Samuel and S.~Watamura,
``Conformal Techniques, Bosonization and Tree Level String Amplitudes,''
  Nucl.\ Phys.\  B {\bf 288}, 173 (1987).
}

\lref\AntoniadisEW{
  I.~Antoniadis,
  ``A Possible new dimension at a few TeV,''
  Phys.\ Lett.\  B {\bf 246}, 377 (1990).
}
\lref\ptmhv{
  S.J.~Parke and T.R.~Taylor,
``An Amplitude for $n$ Gluon Scattering,''
  Phys.\ Rev.\ Lett.\  {\bf 56}, 2459 (1986).
}

\lref\ptsusy{
  S.J.~Parke and T.R.~Taylor,
``Perturbative QCD Utilizing Extended Supersymmetry,''
  Phys.\ Lett.\  B {\bf 157}, 81 (1985)
  [Erratum-ibid.\  {\bf 174B}, 465 (1986)].
}
\lref\BrittoDG{
  R.~Britto, B.~Feng, R.~Roiban, M.~Spradlin and A.~Volovich,
  ``All split helicity tree-level gluon amplitudes,''
  Phys.\ Rev.\  D {\bf 71}, 105017 (2005)
  [arXiv:hep-th/0503198].
}
\lref\LuoRX{
  M. Luo and C. Wen,
``Recursion relations for tree amplitudes in super gauge theories,''
  JHEP {\bf 0503}, 004 (2005)
  [arXiv:hep-th/0501121].
}
\lref\BidderIN{
  S.J.~Bidder, D.C.~Dunbar and W.B.~Perkins,
``Supersymmetric Ward identities and NMHV amplitudes involving gluinos,''
  JHEP {\bf 0508}, 055 (2005)
  [arXiv:hep-th/0505249].
}

\lref\ptfour{
  S.J.~Parke and T.R.~Taylor,
``Gluonic Two Goes to Four,''
  Nucl.\ Phys.\  B {\bf 269}, 410 (1986).
}

\lref\Lykken{J.D.~Lykken,
``Weak Scale Superstrings,''
  Phys.\ Rev.\  D {\bf 54}, 3693 (1996)
  [arXiv:hep-th/9603133];\br
I.~Antoniadis, N.~Arkani-Hamed, S.~Dimopoulos and G.R.~Dvali,
``New dimensions at a millimeter to a Fermi and superstrings at a TeV,''
  Phys.\ Lett.\  B {\bf 436}, 257 (1998)
  [arXiv:hep-ph/9804398].
}

\lref\ManganoBY{
  M.L.~Mangano and S.J.~Parke,
  ``Multiparton amplitudes in gauge theories,''
  Phys.\ Rept.\  {\bf 200}, 301 (1991)
  [arXiv:hep-th/0509223].
}
\lref\DixonWI{
  L.J.~Dixon,
  ``Calculating scattering amplitudes efficiently,''
  arXiv:hep-ph/9601359.
}

\lref\BerendsME{
  F.A.~Berends and W.T.~Giele,
  ``Recursive Calculations for Processes with n Gluons,''
  Nucl.\ Phys.\  B {\bf 306}, 759 (1988).
}
\lref\Peskin{S.~Cullen, M.~Perelstein and M.E.~Peskin,
``TeV strings and collider probes of large extra dimensions,''
  Phys.\ Rev.\  D {\bf 62}, 055012 (2000)
  [arXiv:hep-ph/0001166].
}

\lref\banksii{T.~Banks and L.J.~Dixon,
``Constraints on String Vacua with Space-Time Supersymmetry,''
  Nucl.\ Phys.\  B {\bf 307}, 93 (1988);\br
S.~Ferrara, D.~L\"ust and S.~Theisen,
 ``World Sheet Versus Spectrum Symmetries In Heterotic And Type II Superstrings,''
  Nucl.\ Phys.\  B {\bf 325}, 501 (1989).
}

\lref\JOE{
J. Polchinski, "String Theory'', Sections 6 \& 12, Cambridge University Press 1998.}

\lref\Dan{D.~Oprisa and S.~Stieberger,
``Six gluon open superstring disk amplitude, multiple hypergeometric  series
and Euler-Zagier sums,''
  arXiv:hep-th/0509042.
}

\lref\CSW{F.~Cachazo, P.~Svrcek and E.~Witten,
``MHV vertices and tree amplitudes in gauge theory,''
  JHEP {\bf 0409}, 006 (2004)
  [arXiv:hep-th/0403047].
}

\lref\STii{
  S.~Stieberger and T.R.~Taylor,
``Multi-gluon scattering in open superstring theory,''
  Phys.\ Rev.\  D {\bf 74}, 126007 (2006)
  [arXiv:hep-th/0609175].
}
\lref\Kosower{
  D.A.~Kosower,
  ``Next-to-maximal helicity violating amplitudes in gauge theory,''
  Phys.\ Rev.\  D {\bf 71}, 045007 (2005)
  [arXiv:hep-th/0406175].
}
\lref\STi{
  S.~Stieberger and T.R.~Taylor,
``Amplitude for N-gluon superstring scattering,''
  Phys.\ Rev.\ Lett.\  {\bf 97}, 211601 (2006)
  [arXiv:hep-th/0607184].
}

\lref\STiii{
S. Stieberger and T.R.~Taylor, work in progress.}

\lref\mman{M.L.~Mangano, S.J.~Parke and Z.~Xu,
  ``Duality and Multi - Gluon Scattering,''
  Nucl.\ Phys.\  B {\bf 298}, 653 (1988).
  }

\Title{\vbox{\rightline{MPP--2007--110}\rightline{AEI--2007--167}
}}
{\vbox{\centerline{Complete Six--Gluon Disk Amplitude}
\bigskip\centerline{ in Superstring Theory}
}}
\smallskip
\centerline{Stephan Stieberger$^{a,b}$\ \ and\ \ Tomasz R. Taylor$^{a,c}$}
\bigskip
\centerline{\it $^a$ Max--Planck--Institut f\"ur Physik}
\centerline{\it Werner--Heisenberg--Institut}
\centerline{\it 80805 M\"unchen, Germany}
\vskip7pt
\centerline{\it $^b$ Max--Planck--Institut f\"ur Gravitationsphysik}
\centerline{\it Albert--Einstein--Institut}
\centerline{\it 14476 Potsdam, Germany}
\vskip7pt
\centerline{\it $^c$ Department of Physics}
\centerline{\it Northeastern University}
\centerline{\it Boston, MA 02115, USA}

\bigskip\bigskip
\centerline{\bf Abstract}
\vskip7pt
\noindent
We evaluate all next-to-maximal helicity violating (NMHV) six-gluon amplitudes
in type I open superstring theory in four dimensions, at the disk level,
to all orders in $\ap$. Although the computation utilizes supersymmetric Ward
identities, the result holds for all compactifications, even for those that
break supersymmetry and is completely model-independent. Together with the
maximally helicity violating (MHV) amplitudes presented in the previous work,
our results provide the complete six-gluon disk amplitude.
\Date{}
\noindent

\goodbreak


\newsec{Introduction}
Multi-gluon scattering amplitudes are important from both theoretical and experimental points of view because they describe the processes underlying hadronic jet production at high energy colliders.
Up to the energies accessible to the existing accelerators, there is an excellent agreement between experimental data and the amplitudes calculated in the framework of perturbative Quantum Chromodynamics (QCD). If in the upcoming Large Hadron Collider (LHC) experiments any discrepancy is discovered between QCD and the observed jet cross sections, it will be interpreted as a signal of new physics beyond the standard model.

Among the extensions of the standard model, superstring theory stands out as one of the boldest ones because it incorporates gravity and covers a huge span of energies, up to the Planck mass. However, the fundamental energy scale is the string mass, which need not necessarily to be as high provided that the Universe contains some large extra  dimensions
\refs{\Lykken,\AntoniadisEW}.
If the string scale is within, or not too far above the range of LHC energies, the effects of Regge excitations may be observable, and a direct experimental proof of superstring theory can be at hand. In particular, the multi-gluon amplitudes will be affected by the so-called $\ap$ corrections and the measurement of the corresponding jet cross sections can reveal some spectacular signals of superstring theory.

In a recent series of papers \refs{\Dan,\STi,\STii,\August}, we developed a formalism for computing $N$-gluon amplitudes at the disk level, {\it i.e.} at the leading order in the string/gauge coupling constant but to all orders in $\ap$.
The most important property of these leading contributions is that they are
completely model-independent. These amplitudes are very robust because they
hold for arbitrary compactifications of superstring theory from ten to four
dimensions, including those that
break supersymmetry.\foot{
The difference between supersymmetric and non-supersymmetric amplitudes appears at the one-loop level. In particular, the presence of non-supersymmetric dimension six
operator Tr$(F_{\mu\nu}F^{\nu\rho}F_{\rho\mu})$, where $F$ is the gauge field strength tensor,  may have some interesting phenomenological consequences \refs{\Peskin,\simmons}.} The formalism combines the use of traditional supersymmetry and helicity techniques together with some elements of the theory of multiple hypergeometric integrals that allow expressing the final results in terms of $(N-3)!$ generalized hypergeometric functions of kinematic invariants.
This is particularly effective when applied to the maximally helicity violating \ptmhv\ (MHV) amplitudes \August. Next-to-maximal helicity violating (NMHV) amplitudes which appear starting from $N=6$ seem to have a more complicated structure. In this work, we study the case of six-gluon NMHV amplitudes. We obtain some relatively simple expressions for NMHV amplitudes which complement the MHV amplitudes presented in \refs{\STi,\STii,\August}, providing the full six-gluon disk amplitude.

The paper is organized as follows. In Section 2, we present  supersymmetry (SUSY) relations that allow expressing all three independent six-gluon NMHV amplitudes in terms of the amplitudes
involving four scalars and two gluons, four scalars and two gauginos, and six
scalars. In Section 3, we evaluate these auxiliary amplitudes and express them in terms
of certain hypergeometric integrals.
In Section 4, we combine them according to SUSY relations and obtain explicit expressions for all NMHV amplitudes. We show that the leading order of the expansion in powers of $\ap$ correctly reproduces the QCD result. In two Appendices, we list the relevant hypergeometric functions and give their $\ap$-expansions up to the next-to-leading order ${\cal O}(\ap^2)$ with respect to the leading (QCD) contributions.

\newsec{SUSY Relations for NMHV Amplitudes}

The full six-gluon NMHV amplitude can be constructed from three {\it
partial\/} subamplitudes \refs{\ManganoBY,\DixonWI}, each associated to the
same Chan-Paton factor ${\rm Tr} \, (\, T^{a_1}\cdots T^{a_6})$, but characterized by three inequivalent helicity orderings:
\eqn\threedef{\eqalign{A^Y\equiv &\ A(g^{-}_1,g^{-}_2, g^{+}_3,g^{+}_4,g^{-}_5,g^{+}_6)\, ,\cr
A^X\equiv &\ A(g^{+}_1,g^{+}_2, g^{-}_3,g^{-}_4,g^{-}_5,g^{+}_6)\, ,\cr A^Z\equiv &\ A(g^{-}_1,g^{+}_2, g^{-}_3,g^{+}_4,g^{-}_5,g^{+}_6)\, .}}
These amplitudes will be expressed in terms of the following amplitudes with the gluons replaced by scalars or fermions:
\eqn\threeamps{\eqalign{A^Y_{\lambda}\equiv  A(\phi^{-}_1,\phi^{-}_2, \phi^{+}_3,\phi^{+}_4,\lambda^{-}_5,\lambda^{+}_6)&~,\qquad A^Y_{s}\equiv A(\phi^{-}_1,\phi^{-}_2, \phi^{+}_3,\phi^{+}_4,\phi^{-}_5,\phi^{+}_6)\, ,\cr
A^X_{\lambda}\equiv  A(\phi^{+}_1,\phi^{+}_2, \phi^{-}_3,\phi^{-}_4,\lambda^{-}_5,\lambda^{+}_6)&~,\qquad A^X_{s}\equiv A(\phi^{+}_1,\phi^{+}_2, \phi^{-}_3,\phi^{-}_4,\phi^{-}_5,\phi^{+}_6)\, ,\cr A^Z_{\lambda}\equiv A(\phi^{-}_1,\phi^{+}_2, \phi^{-}_3,\phi^{+}_4,\lambda^{-}_5,\lambda^{+}_6)&~,\qquad A^Z_{s}\equiv A(\phi^{-}_1,\phi^{+}_2, \phi^{-}_3,\phi^{+}_4,\phi^{-}_5,\phi^{+}_6)\, ,}}
and the  amplitudes with four gluons replaced by scalars:
\eqn\threegs{\eqalign{A^Y_g\equiv &\ A(\phi^{-}_1,\phi^{-}_2, \phi^{+}_3,\phi^{+}_4,g^{-}_5,g^{+}_6)\, ,\cr
A^X_g\equiv &\ A(\phi^{+}_1,\phi^{+}_2, \phi^{-}_3,\phi^{-}_4,g^{-}_5,g^{+}_6)\, ,\cr A^Z_g\equiv &\ A(\phi^{-}_1,\phi^{+}_2, \phi^{-}_3,\phi^{+}_4,g^{-}_5,g^{+}_6)\, .}}
Here, $\phi$ is the scalar component of $\Nc=2$ gauge supermultiplet and $\lambda$ is one of the two gauginos \August.

Recently, we showed \August\ that all field-theoretical SUSY relations
\refs{\Grisaru,\ptsusy} between scattering amplitudes hold also in superstring theory at the disk level, to all orders in $\ap$. Actually, we find  most useful the relations already used in the original computation of the six-gluon QCD amplitudes \ptfour. In order to write these relations down, we first introduce the following  kinematic variables
\eqn\kin{\eqalign{Y=&\ k_3+k_4+k_6~,\qquad \alpha_Y=-\,\langle 12\rangle[34][\,6|Y|5\rangle~,\qquad y=\langle 12\rangle[34]\,Y^2~,\cr
X=&\ k_1+k_2+k_6~,\qquad \alpha_X=-[12]\langle 34\rangle[\,6|X|5\rangle~,\qquad x=[12]\langle 34\rangle X^2~,\cr
Z=&\ k_2+k_4+k_6~,\qquad \alpha_Z=-\,\langle 13\rangle [24][\,6|Z\,|5\rangle~,\qquad z=\langle 13\rangle [24]\, Z^2~,}}
depending on the momenta $k_1,k_2,\dots ,k_6$. Here, we used the standard
notation
\refs{\ManganoBY,\DixonWI} for spinor products, in particular:
\eqn\exam{[\,6|Y|5\rangle=[63 ]\langle\, 3 5\rangle+[6 4]\langle 45\rangle~,\quad {\it etc.}}
{}For the scalar products of momenta, we use the notation of \STii:
\eqn\conv{s_{ij}=2\ap k_ik_j~,\quad s_i=\ap(k_i+k_{i+1})^2~,\quad t_i=\ap(k_i+k_{i+1}+k_{i+2})^2 \qquad     (i+6\equiv i)~.}
All scalar products $s_{ij}$ can be expressed in terms of $s_1,s_2,s_3,s_4,s_5,s_6$ and $t_1,t_2,t_3$ \STii, {\it e.g}. $s_{12}=s_1$ {\it etc}. Now the SUSY relations can be written as
\eqn\rels{\eqalign{A^Y= &\ ~{\ap^4\over s_{12}^2s_{34}^2}\big(\,y^2 A^Y_g- 2\,y\,\alpha_Y A^Y_{\lambda}+\alpha_Y^2 A^Y_s\big)~ ,\cr A^X= &\ ~{\ap^4\over s_{12}^2s_{34}^2}\big(\,x^2 A^X_g- 2\,x\,\alpha_X A^X_{\lambda}+\alpha_X^2 A^X_s\big)~,
\cr A^Z= &\ ~{\ap^4\over s_{13}^2s_{24}^2}\big(\,z^2 A^Z_g- 2\,z\,\alpha_Z A^Z_{\lambda}+\,\alpha_Z^2 A^Z_s\,\big)~, }}
where the factors $\ap^4$ appear artificially, due to the choice of string units in \conv.
Note that the latter two relations ($X,Z$) follow from the first one ($Y$) by the replacements $(1\leftrightarrow 4,2\leftrightarrow 3)$ and $(2\leftrightarrow 3)$, respectively.
In the next Section, we  evaluate the amplitudes appearing on the r.h.s.\ of
Eq. \rels.

\newsec{Six--point disk scattering of scalars, gauginos and vectors}

In this Section, we compute the six--point string amplitudes \threeamps\ and \threegs\
involving scalars, gauginos and vectors of the four-dimensional $\Nc=2$ vector multiplet.
The world--sheet of the string S--matrix is described by a disk with
all external states $\Phi^a$ created by vertex operators $V_{\Phi^a}$
at the boundary the disk.
In the notation of Refs. \refs{\STi,\STii,\August},
the partial amplitude associated to  the ${\rm Tr}  (\, T^{a_1}\cdots T^{a_6})$ Chan-Paton
factor takes the form:
\eqn\Sechs{\eqalign{
A(\Phi^{a_1},\Phi^{a_2},\Phi^{a_3},&\Phi^{a_4},\Phi^{a_4},\Phi^{a_6})=
V_{CKG}^{-1}\int\limits_{z_1<\ldots<z_6}
\lf(\prod_{k=1}^6 dz_k\ri)\cr
&\times\vev{V_{\Phi^{a_1}}(z_1)\ V_{\Phi^{a_2}}(z_2)\
V_{\Phi^{a_3}}(z_3)\ V_{\Phi^{a_4}}(z_4)\ V_{\Phi^{a_5}}(z_5)\ V_{\Phi^{a_6}}(z_6)}\ .}}
In order to cancel the total background ghost charge in the disk
correlator \Sechs, the vertex operators have to be chosen
in the appropriate ghost picture.
Furthermore, in Eq. \Sechs, the factor $V_{CKG}$ accounts for the volume of the
conformal Killing group of the disk after choosing the conformal gauge.
It will be canceled by fixing three vertex positions and introducing the
respective $c$--ghost correlator.
The gaugino vertex operators, in the $(-1/2)$-ghost picture,
are:\foot{The open string vertex couplings are $g_\phi=(2\ap)^{1/2}\ g_{Y\! M}$,\
$g_\lambda=(2\ap)^{1/2}\ap^{1/4}\ g_{Y\! M}$, and $g_A=(2\ap)^{1/2}\ g_{Y\! M}$
for the scalar, gaugino and vector, respectively.
The $D=4$ gauge coupling $g_{Y\! M}$ can be expressed in terms of
the ten--dimensional gauge coupling $g_{10}$
and the dilaton field $\phi_{10}$ through the relation $g_{Y\! M}=g_{10}e^{\phi_{10}/2}$
\JOE.}
\eqn\gaugino{\eqalign{
V_{\lambda^{a,I}}^{(-1/2)}(z,u,k)
&=g_\lambda\ T^a\ e^{-\phi/2}\ u^\al\ S_\al\ \Si^I\ e^{ik_\rho X^\rho}\ ,\cr
V_{\ov\lambda^{a,I}}^{(-1/2)}(z,u,k)
&=g_\lambda\ T^a\ e^{-\phi/2}\ \ov u_{\dot\bet}\ S^{\dot\bet}\ \ov\Si^I\
e^{ik_\rho X^\rho}\ \ \ ,\ \ \ I=1,2\ .}}
Here $S_\al,S_{\dot\al}$ are the spin fields with the
indices $\al$ (or $\dot\al$)  denoting negative (positive) chirality
in four dimensions. Furthermore, $\phi$ is the scalar bosonizing the superghost system.
In the above definitions, $T^a$ are the Chan--Paton factors accounting for the gauge
degrees of freedom of the two open string ends.
The on--shell constraints $k^2=0,\ \slashchar{k}u=0$ are imposed.

For $\Nc=2$ space--time SUSY the internal SCFT splits into two pieces.
One piece is the $c=3$ superconformal algebra, which corresponds
to a torus compactificaton with the two
complex internal fermions $\Psi^{\mp}=e^{\pm iH_3}$. The second piece
represents a $c=6$ superconformal algebra, which
contains the $SU(2)$ currents $J_3=i\p H,\ J^{12}=e^{i\sqrt 2 H}$
and $J^{21}=e^{-i\sqrt 2 H}$ \banksii. The (internal) Ramond fields $\Si^I$ may be
expressed by these bosonic fields $H_3$ and $H$:
\eqn\current{
\Si^1=e^{\fc{i}{2}H_3}\ e^{\fc{i}{\sqrt2}H}\ \ \ ,\ \ \ \Si^2=e^{\fc{i}{2}H_3}\
e^{-\fc{i}{\sqrt2}H}\ .}

Finally, the vertex operators $V_{\phi^{a,\pm}}(z,k)$
for the scalars and for the vectors $V_{A^a}(z,\xi,k)$
can be found in Section 3 of \August.

The six--point correlator in the integrand of \Sechs\ is evaluated by performing
all possible Wick contractions. All three and four--point fermionic correlators
involving fermions and  spin fields are given in
\threeref\CFT\KLLSW\August, while an important five--point correlator will be
computed below.
Because of the $PSL(2,\IR)$ invariance on the disk, we can fix three positions
of the vertex operators. A convenient choice respecting the
ordering $z_1<\ldots<z_6$ is
\eqn\choice{
z_1=-z_\infty=-\infty\ \ \ ,\ \ \ z_2=0\ \ \ ,\ \ \ z_3=1\ ,}
which implies the ghost factor $\vev{c(z_1)c(z_2)c(z_3)}=-z_\infty^2$.
The remaining three vertex positions take arbitrary values inside
the integration domain $1<z_4<z_5<z_6<~\infty$. The latter is parameterized by
$z_4=x^{-1},\ z_5=(xy)^{-1}$ and $z_6=(xyz)^{-1}$, with $0<x,y,z<1$.
Generally for this choice, the integrand of \Sechs\ contains the
common factor \doubref\Dan\STii
\eqn\Integrand{\eqalign{
\Ic(x,y,z)&=x^{s_2}\ y^{t_2}\ z^{s_6}\
(1-x)^{s_3}\ (1-y)^{s_4}\ (1-z)^{s_5}\cr
&\times(1-xy)^{t_3-s_3-s_4}\ (1-yz)^{t_1-s_4-s_5}\ (1-xyz)^{s_1+s_4-t_1-t_3}\ .}}
The resulting integrals represent generalized Euler integrals and integrate to
multiple Gaussian hypergeometric functions \refs{\Dan,\STii}.

In order to correctly normalize the amplitudes, some additional factors
have to be
taken into account.  They stem from determinants
and Jacobians of certain path integrals. On the disk, the net result of those
contributions is an additional factor of $C_{D_2}={1\over 2\, g_{Y\! M}^2\,\ap^2}$
which must be included in all disk correlators \JOE.

\subsec{Four scalars and two gauginos}

\noindent
First, we compute the three subamplitudes \threeamps\ involving four scalars and two gauginos: $A^Y_{\lambda}$, $A^X_{\lambda}$ and $A^Z_{\lambda}$. To that end, we evaluate the amplitude \Sechs\ with the following correlator:
\eqn\SSSSGG{
\vev{V_{\phi^{a_1}}^{(0)}(z_1,k_1)V_{\phi^{a_2}}^{(0)}(z_2,k_2)
V_{\phi^{a_3}}^{(0)}(z_3,k_3)V_{\phi^{a_4}}^{(-1)}(z_4,k_4)
V^{(-1/2)}_{\lambda^{a_5,I}}(z_5,u_5,k_5)
V^{(-1/2)}_{\ov\lambda^{a_6,J}}(z_6,\ov u_6,k_6)}\ ,}
{}for the helicity configurations $Y$, $X$ and $Z$.

To compute \SSSSGG, we need\foot{Throughout this article we adapt to the
notation and spinor
algebra of the book of Wess and Bagger. In particular, spinor indices are raised and
lowered with the anti--symmetric tensors $\eps_{\al\bet}$ and $\eps^{\dot\al\dot\bet}$.
Besides spinor products are defined to be
$\chi\eta=\chi^\al\eps_{\al\bet}\eta^\bet$ ($\ov\chi\ov\eta=\ov\chi_{\dot\al}
\eps^{\dot\al\dot\bet}\ov\eta_{\dot\bet}$) for some spinors
$\chi,\eta$ ($\ov\chi,\ov\eta$).} the five--point function
\eqn\psiSS{\eqalign{
&\vev{\psi^{\lambda_1}(z_1)\psi^{\lambda_2}(z_2)\psi^{\lambda_3}(z_3)S_\al(z_4)
S_{\dot\bet}(z_5)}=\fc{1}{\sqrt 2}\ (z_{14}z_{15}z_{24}z_{25}z_{34}z_{35})^{-1/2}\cr
&\hskip0.4cm\times\lf\{\fc{z_{24}z_{35}}{z_{23}}\
\si^{\lambda_1}_{\al\dot\bet}\
\delta^{\lambda_2\lambda_3}-
\fc{z_{14}z_{35}}{z_{13}}\ \si^{\lambda_2}_{\al\dot\bet}\
\delta^{\lambda_1\lambda_3}+\fc{z_{14}z_{25}}{z_{12}}\
\si^{\lambda_3}_{\al\dot\bet}\
\delta^{\lambda_1\lambda_2}+\h\ z_{45}\
(\si^{\lambda_1}\ov\si^{\lambda_2}\si^{\lambda_3})_{\al\dot\bet}\ri\},}}
which may be derived by studying its singular behavior and by using equations written in
\August. The last term in \psiSS\ may also be rewritten thanks to the
identity\foot{With this identity the correlator \psiSS\ becomes:
$$\eqalign{
&\vev{\psi^{\lambda_1}(z_1)\psi^{\lambda_2}(z_2)\psi^{\lambda_3}(z_3)S_\al(z_4)
S_{\dot\bet}(z_5)}=\fc{1}{2\sqrt 2}\
(z_{14}z_{15}z_{24}z_{25}z_{34}z_{35})^{-1/2}\ \{ i\
\eps^{\lambda_1\lambda_2\lambda_3\lambda}
\ (\si_{\lambda})_{\al\dot\bet}\ z_{45}\cr
&+\fc{\si^{\lambda_1}_{\al\dot\bet}\ \delta^{\lambda_2\lambda_3}}{z_{23}}\
(z_{24}z_{35}+z_{25}z_{34})-
\fc{\si^{\lambda_2}_{\al\dot\bet}\ \delta^{\lambda_1\lambda_3}}{z_{13}}\
(z_{14}z_{35}+z_{15}z_{34})+\fc{\si^{\lambda_3}_{\al\dot\bet}\
\delta^{\lambda_1\lambda_2}}{z_{12}}\ (z_{14}z_{25}+z_{15}z_{24})\ \}\ .}$$}:
$i\eps^{\lambda_1\lambda_2\lambda_3\lambda}\si_{\lambda}=
\si^{\lambda_1}\ov\si^{\lambda_2}\si^{\lambda_3}+\delta^{\lambda_2\lambda_3}
\si^{\lambda_1}-\delta^{\lambda_1\lambda_3}\si^{\lambda_2}+
\delta^{\lambda_1\lambda_2}\si^{\lambda_3}$.
Furthermore, we need the two correlators of internal fields:
\eqn\internal{\eqalign{
\vev{\Psi(z_1)\ov\Psi(z_2)\Si^I(z_3)\ov\Si^J(z_4)}&=
\delta^{IJ}\ z_{12}^{-1}\ z_{34}^{-3/4}\
\lf(\fc{z_{13}z_{24}}{z_{14}z_{23}}\ri)^{1/2}\ ,\cr
\vev{\Psi(z_1)\Psi(z_2)\ov\Psi(z_3)\ov\Psi(z_4)\Si^I(z_5)\ov\Si^J(z_6)}&=\delta^{IJ}\
\fc{z_{12}z_{34}}{z_{13}z_{14}z_{23}z_{24}}\
\lf(\fc{z_{15}z_{25}z_{36}z_{46}}{z_{16}
z_{26}z_{35}z_{45}}\ri)^{1/2}\ z_{56}^{-3/4}\ .}}
All remaining correlators appearing in \SSSSGG\ are basic and can be found in  \August.

After assembling everything in the correlators \SSSSGG,
each of the partial amplitudes $Y$, $X$ and $Z$ takes the form
\eqn\ssssgg{
A_\lambda=4\ap^2\ \gym^4\  \big(\, [\, 6| 1| 5\rangle L_1
+[\, 6| 2| 5\rangle L_2+[\, 6| 3| 5\rangle L_3-
\ap[\, 6| 3| 2\rangle[ 2| 1| 5\rangle L_4\, \big)\ ,}
with the set of four functions
$L_i\in\{L^Y_i,L^X_i,L_i^Z\},~ i=1,2,3,4,$ specific to the three helicity
configurations $Y$, $X$ and $Z$, respectively.
The integral representations of these
functions are given in Appendix \appA. Actually, the kinematic factor in front
of $L_4$ can be expressed in terms of those in front of $L_{1,2,3}$, and the
result \ssssgg\ can be simplified to
\eqn\sssshh{
A_\lambda=4\ap^2\ \gym^4\  \big(\, [\, 6| 1| 5\rangle\, H_1
+[\, 6| 2| 5\rangle\, H_2+[\, 6| 3| 5\rangle\, H_3 \,\big)\ ,}
where:
\eqn\LPX{\eqalign{
H_1&=L_1-\fc{L_4}{2s_5}\ (s_{25}s_{36}-s_{26}s_{35}+s_{23}s_{56})\ ,\cr
H_2&=L_2-\fc{L_4}{2s_5}\ (s_{16}s_{35}-s_{15}s_{36}-s_{13}s_{56})\ ,\cr
H_3&=L_3-\fc{L_4}{2s_5}\ (s_{15}s_{26}-s_{16}s_{25}+s_{12}s_{56})\ .}}

\subsec{Six scalars}

\noindent
Here, we compute the three six-scalar subamplitudes \threeamps: $A^Y_{s}$,
$A^X_{s}$ and $A^Z_{s}$. To that end, we evaluate the amplitude \Sechs\ with the  correlators
\eqn\SSSSSS{
\vev{V_{\phi^{a_1}}^{(0)}(z_1,k_1)V_{\phi^{a_2}}^{(0)}(z_2,k_2)
V_{\phi^{a_3}}^{(0)}(z_3,k_3)V_{\phi^{a_4}}^{(0)}(z_4,k_4)
V^{(-1)}_{\phi^{a_5}}(z_5,k_5)V^{(-1)}_{\phi^{a_6}}(z_6,k_6)}\ ,}
{}for the helicity configurations $Y$, $X$ and $Z$.
After a straightforward calculation, we obtain
\eqn\ssssss{
A_s=4\ap\ \gym^4\ L_5\ ,}
with the functions $L_5\in\{L_5^Y,L_5^X,L_5^Z\}$ specific to the three
helicity configurations $Y$, $X$ and $Z$, respectively.
Again, we present the integrals $L_5$ for the three cases in Appendix \appA.

\subsec{Four scalars and two vectors}

\noindent
Finally, we evaluate the three subamplitudes \threegs\ involving four scalars
and two gauge fields: $A^Y_{g}$, $A^X_{g}$ and $A^Z_{g}$. To that end, we
evaluate the amplitude \Sechs\ with the  correlators
\eqn\SSSSSS{
\vev{V_{\phi^{a_1}}^{(0)}(z_1,k_1)V_{\phi^{a_2}}^{(0)}(z_2,k_2)
V_{\phi^{a_3}}^{(0)}(z_3,k_3)V_{\phi^{a_4}}^{(0)}(z_4,k_4)
V^{(-1)}_{A^{a_5}}(z_5,\xi_5,k_5)V^{(-1)}_{A^{a_6}}(z_6,\xi_6,k_6)}\ ,}
{}for the helicity configurations $Y$, $X$ and $Z$.
The  amplitude $A^Y_g$ has already been computed in \August.
In fact, all three (partial) amplitudes have a similar form:
\eqn\Becomes{\eqalign{
A_g=&\ 8\ap^2\ g_{Y\! M}^4
\lf[\ (\xi_5k_3)(\xi_6k_2)\ K_1+(\xi_5k_2)(\xi_6k_3)\
K_2+(\xi_5k_1)(\xi_6k_2)\ K_3\ri.\cr
&+(\xi_5k_1)(\xi_6k_3)\ K_4+(\xi_5k_2)(\xi_6k_1)\ K_5+(\xi_5k_3)(\xi_6k_1)\ K_6  \cr
&+(\xi_5k_3)(\xi_6k_4)\ K_7
+(\xi_5k_4)(\xi_6k_3)\ K_8 +(\xi_5k_2)(\xi_6k_4)\ K_9\cr
&\lf. +(\xi_5k_4)(\xi_6k_2)\ K_{10}+(\xi_5k_1)(\xi_6k_4)\ K_{11}+(\xi_5k_4)(\xi_6k_1)\ K_{12}+
\,(\xi_5\xi_6)\ K_{13}\ \ri]\ ,}}
specified by thirteen functions $K_i$ for each configuration $Y$, $X$ and $Z$.
Here, $\xi_5$ and $\xi_6$ are the gluon polarization vectors. Actually, one also finds
$A_g^Y=A_g^X$, therefore $K_i^Y=K_i^X$.
In order to write Eq. \Becomes\ more explicitly, we choose $k_6$ as the
reference
vector for
the (negative) polarization vector of gluon $g_5$ and $k_5$ as the reference vector for
the (positive) polarization vector of gluon $g_6$. Then
\eqn\epsk{ \xi^-_5\xi_6^+=0    \quad,\quad
(\xi^-_5k_i)(\xi^+_6k_j)=-\ap{[\, 6\, i]\langle i 5\rangle[\,6j] \langle j5\rangle \over 2s_5}~,}
and the amplitude \Becomes\ can be rewritten as
\eqn\agg{\eqalign{
A_g= -{4\,\ap^3 \ g_{Y\! M}^4 \over s_5}&
\lf(\  [\, 6| 2| 5\rangle\ [\,6|3|5\rangle\, G_1+
[\, 6| 1| 5\rangle\ [\,6|2|5\rangle\, G_2+[\, 6| 1| 5\rangle\ [\,6|3|5\rangle\, G_3
\ri.   \cr
&\lf.+[\, 6| 3| 5\rangle\ [\,6|4|5\rangle\, G_4+
[\, 6| 2| 5\rangle\ [\,6|4|5\rangle\, G_5+
[\, 6| 1| 5\rangle\ [\,6|4|5\rangle\, G_6\ \ri)\ ,}}
where $G_1=K_1+K_2,\ G_2=K_3+K_5,\ G_3=K_4+K_6,\ G_4=K_7+K_8,\
G_5=K_9+K_{10}$ and $G_6=K_{11}+K_{12}$.
The integral representations of all these functions are given in Appendix~\appA.

\newsec{Six-gluon NMHV Amplitudes}

After computing all auxiliary amplitudes, we are now in a position to write
down the six-gluon amplitudes. The result is obtained by substituting
Eqs.\sssshh,
\ssssss\ and \agg\ into the r.h.s.\ of SUSY relations \rels. We could leave
this result as it is, however there are at least two good reasons for trying
to combine all contributions into a more compact form. First, although for each
helicity configuration, the
amplitude depends on ten functions $G$, $H$ and $L_5$, we
know that  only $(N-3)!=6$ of them are independent \refs{\STi,\STii,\Dan}.
Indeed, by using techniques developed in \Dan, it is possible to find
relations between these functions. In what follows, we will combine the ten integrals
$G$, $H$ and $L_5$ into a set of six functions $N_i$, for each helicity
configuration.
Second, the kinematic factors appearing in auxiliary amplitudes
are related, therefore they can be combined to a form involving fewer
kinematic factors.
This is highly desirable for many reasons, especially for the comparison of
the $\ap=0$ limit with the well-known QCD amplitudes.
The six functions $N_i$ will appear naturally in this context.
In QCD, six-gluon NMHV amplitudes were first calculated in \ptfour,
and later recast in an elegant form in
Refs.\refs{\mman,\ManganoBY}.\foot{For more recent work on NMHV amplitudes,
see {\it e.g}.\ Refs. \refs{\Kosower,\BrittoAP,\BrittoDG,\LuoRX,\BidderIN}.}

The basic relations that allow combining various contributions on the r.h.s.\
of Eq. \rels\ are:
\eqn\basic{Y^2[16]+[15][\,6|Y|5\rangle=-[12][\,6|Y|2\rangle~,}
and its variations obtained by applying various permutations of
$\{1,2,3,4,5,6\}$
and/or complex conjugation. Eq. \basic\ follows from Schouten's identity and
momentum conservation. The goal is to rewrite our results in a form similar to
QCD amplitudes collected in Eqs.(5.28) and Table 4 of Ref. \ManganoBY. Below,
we list the amplitudes obtained by manipulating kinematic factors on the
r.h.s.\ of Eq. \rels, for the three helicity configurations separately.

\subsec{$Y$-configuration and its $\ap=0$ limit}

The result will be expressed in terms of the following kinematic variables:
\eqn\abcy{  \alpha_Y=-\,\langle 12\rangle[34][\,6|Y|5\rangle~,\qquad
\beta_Y=\langle 12\rangle[46][\,3|Y|5\rangle~,\qquad
\gamma_Y=\langle 51\rangle[34][\,6|Y|2\rangle~,}
where $\alpha_Y$, already defined in Eq. \kin, is listed for completeness. Then
\eqn\agy{\eqalign{
 A(& g^{-}_1,g^{-}_2, g^{+}_3,g^{+}_4,g^{-}_5,g^{+}_6) = A^Y=
{\rm Tr}(\,T^{a_1}T^{a_2}T^{a_3}T^{a_4}T^{a_5}T^{a_6})\ {(\sqrt{2}\,g_{Y\!
M})^4\,\ap^5  \over s_5}~
\cr &\ \times \lf(\,N^Y_1{\alpha_Y^2\over s_1^2s_3^2} ~+ N^Y_2{\beta_Y^2\over
s_1^2} ~+N^Y_3{\gamma_Y^2\over s_3^2} ~+ N^Y_4{\alpha_Y\beta_Y\over s_1^2s_3}
~+ N^Y_5{\alpha_Y\gamma_Y\over s_1s_3^2} ~+N^Y_6{\beta_Y\gamma_Y\over
s_1s_3}\, \ri)\ ,}}
with the functions $N^Y$ written below:
\eqn\NY{\eqalign{
N^Y_1&=-(s_1+s_3+s_6-t_2-t_3)\ (s_1+s_3+s_4-t_1-t_3)\ G^Y_1
-2\ s_5\ (s_5+s_6-t_2)\ H_1^Y\cr
&-(s_1+s_3+s_6-t_2-t_3)\ [\ (s_5+s_6-t_2)\ G^Y_2-(s_4+s_5-t_1)\ G^Y_5-2\ s_5\ H_2^Y\ ]\cr
&+(s_1+s_3+s_4-t_1-t_3)\ [\ (s_5+s_6-t_2)\ G^Y_3-(s_4+s_5-t_1)\ G^Y_4-2\ s_5\ H_3^Y\ ]\cr
&-(s_4+s_5-t_1)\ (s_5+s_6-t_2)\ G_6^Y-s_5\ L^Y_5\ ,\cr
N^Y_2&=-G^Y_4\ ,\cr
N^Y_3&=-G^Y_2\ ,\cr
N^Y_4&=(s_1+s_3+s_6-t_2-t_3)\ (G^Y_1-G^Y_5)+(s_1+s_3+2\ s_4+s_5-2\ t_1-t_3)\
G^Y_4\cr
&-(s_5+s_6-t_2)\ (G^Y_3-G^Y_6)+2\ s_5\ H_3^Y\ ,\cr
N^Y_5&=(s_1+s_3+s_4-t_1-t_3)\ (G^Y_1-G^Y_3)+(s_1+s_3+s_5+2\ s_6-2\ t_2-t_3)\
G^Y_2\cr
&-(s_4+s_5-t_1)\ (G^Y_5-G^Y_6)+2\ s_5\ (H_1^Y-H_2^Y)\ ,\cr
N^Y_6&=-G^Y_1+G^Y_3+G^Y_5-G^Y_6\ .}}
Their low-energy expansions are
\eqn\ny{\eqalign{
N^Y_1&=\fc{s_1s_3s_5}{s_4s_6t_3}-\zeta(2)\
\lf(s_1s_3-\fc{s_1s_3t_1}{s_4}-\fc{s_1s_3t_2}{s_6}+\fc{s_1s_2s_3s_5}{s_4s_6}
+\fc{s_1^2s_3s_5}{s_4t_3}+\fc{s_1s_3^2s_5}{s_6t_3}\ri)+\ldots\ ,\cr
N^Y_2&=\fc{s_1}{s_2s_4t_1}-\zeta(2)\
\lf(\fc{s_1s_6}{s_2s_4}+\fc{s_1^2}{s_4t_1}+\fc{s_1s_5}{s_2t_1}\ri)+\ldots\ ,\cr
N^Y_3&=\fc{s_3}{s_2s_6t_2}-\zeta(2)\
\lf(\fc{s_3s_4}{s_2s_6}+\fc{s_3s_5}{s_2t_2}+\fc{s_3^2}{s_6t_2}\ri)+\ldots\ ,\cr
N^Y_4&=\fc{s_1t_2}{s_2s_4s_6}-\zeta(2)\ \lf(\fc{s_1(s_1-s_3-s_5-t_3)}{s_4}-
\fc{s_1t_2}{s_2}+\fc{s_1t_1t_2}{s_2s_4}+\fc{s_1t_2^2}{s_2s_6}+\fc{s_1t_2t_3}{s_4s_6}\ri)+
\ldots\ ,\cr
N^Y_5&=\fc{s_3t_1}{s_2s_4s_6}+\zeta(2)\ \lf(\fc{s_3(s_1-s_3+s_5+t_3)}{s_6}+
\fc{s_3t_1}{s_2}-\fc{s_3t_1^2}{s_2s_4}-\fc{s_3t_1t_2}{s_2s_6}-\fc{s_3t_1t_3}{s_4s_6}\ri)+
\ldots\ ,\cr
N^Y_6&=\fc{t_3}{s_2s_4s_6}+\zeta(2)\ \lf(\fc{s_1+s_3-s_5}{s_2}-\fc{t_1t_3}{s_2s_4}-
\fc{t_2t_3}{s_2s_6}-\fc{t_3^2}{s_4s_6}\ri)+\ldots\ ,}}
where dots represent terms suppressed by a factor of order ${\cal
O}\big(\zeta(3)\ap^3\big)$ with respect to the leading term.
The above expansions have been obtained by collecting the expansions of $G$ and $L$ functions
listed in Appendix A.

In the $\ap=0$ limit of the amplitude \agy, only the leading terms of Eq.\ny\
survive.
Then all $\ap$ factors cancel and Eq. \agy\ agrees with the QCD amplitude
written in Eq.(5.28) and Table 4 of Ref.\ManganoBY.\foot{In order to compare, a cyclic
permutation $\{1,2,3,4,5,6\}\to\{3,4,5,6,1,2\}$ must be performed on the
result of Ref. \ManganoBY.}

\subsec{$X$-configuration and its $\ap=0$ limit}

The result will be expressed in terms of the following kinematic variables:
\eqn\abcx{  \alpha_X=-\,[12]\langle 34\rangle[\,6|X|5\rangle~,\qquad
\beta_X=[12]\langle 45\rangle[\,6|X|3\rangle~,\qquad
\gamma_X=[\,61]\langle 34\rangle[\,2|X|5\rangle~,}
where $\alpha_X$, already defined in Eq. \kin, is listed for completeness. Then
\eqn\agx{\eqalign{ A(& g^{+}_1,g^{+}_2, g^{-}_3,g^{-}_4,g^{-}_5,g^{+}_6)   =A^X
={\rm Tr}(\,T^{a_1}T^{a_2}T^{a_3}T^{a_4}T^{a_5}T^{a_6})\
{(\sqrt{2}\,g_{Y\! M})^4\,\ap^5  \over s_5}\cr
&\ \times \lf(\,N^X_1{\alpha_X^2\over s_1^2s_3^2} ~+ N^X_2{\beta_X^2\over
s_1^2} ~+N^X_3{\gamma_X^2\over s_3^2} ~+ N^X_4{\alpha_X\beta_X\over s_1^2s_3}
~+ N^X_5{\alpha_X\gamma_X\over s_1s_3^2} ~+N^X_6{\beta_X\gamma_X\over
s_1s_3}\, \ri)\ ,}}
with the functions $N^X$ written below:
\eqn\NX{\eqalign{
N^X_1&=-(s_4-t_3)\ (s_6-t_3)\ G^X_1-(s_6-t_3)\ (s_6\ G^X_2-s_4\ G_5^X-2\ s_5\ H_2^X)\cr
&+(s_4-t_3)\ (s_6\ G_3^X-s_4\ G_4^X-2\ s_5\ H^X_3)-s_6\ (s_4\ G_6^X+2\ s_5\
H_1^X)-s_5\ L^X_5\ ,\cr
N^X_2&=-G^X_4\ ,\cr
N^X_3&=-G^X_2\ ,\cr
N^X_4&=-(s_6-t_3)\ (G^X_1-G^X_5)+s_6\ (G^X_3-G^X_6)-(2\ s_4-t_3)\ G^X_4-2\ s_5\
H^X_3\ ,\cr
N^X_5&=-(s_4-t_3)\ (G^X_1-G^X_3)-(2\ s_6-t_3)\ G^X_2+s_4\ (G^X_5-G^X_6)-
2\ s_5\ (H^X_1-H^X_2)\ ,\cr
N^X_6&=-G^X_1+G^X_3+G^X_5-G^X_6\ .}}
Their low-energy expansions are
\eqn\nx{\eqalign{
N^X_1&=-\zeta(2)\ s_1s_3+\ldots\ ,\cr
N^X_2&=\fc{s_1}{s_2s_4t_1}-\zeta(2)\
\lf(\fc{s_1s_6}{s_2s_4}+\fc{s_1^2}{s_4t_1}+\fc{s_1s_5}{s_2t_1}\ri)+\ldots\ ,\cr
N^X_3&=\fc{s_3}{s_2s_6t_2}-\zeta(2)\
\lf(\fc{s_3s_4}{s_2s_6}+\fc{s_3s_5}{s_2t_2}+\fc{s_3^2}{s_6t_2}\ri)+\ldots\ ,\cr
N^X_4&=\zeta(2)\ \lf(\fc{s_1t_2}{s_2}+\fc{s_1t_3}{s_4}\ \ri)+\ldots\ ,\cr
N^X_5&=\zeta(2)\ \lf(\fc{s_3t_1}{s_2}+\fc{s_3t_3}{s_6}\ \ri)+\ldots\ ,\cr
N^X_6&=\fc{t_3}{s_2s_4s_6}+\zeta(2)\ \lf(\fc{s_1+s_3-s_5}{s_2}-\fc{t_1t_3}{s_2s_4}-
\fc{t_2t_3}{s_2s_6}-\fc{t_3^2}{s_4s_6}\ri)+\ldots\ ,}}
where dots represent terms suppressed by a factor of order
${\cal O}\big(\zeta(3)\ap^3\big)$ with respect to the leading (QCD)
contribution. The $\ap=0$ limit of the amplitude \agx\ agrees with
Ref.\ManganoBY.\foot{In this case, a cyclic permutation
$\{1,2,3,4,5,6\}\to\{6,1,2,3,4,5\}$ must be performed on the result of
Ref. \ManganoBY.}
Note, in particular, that all terms multiplying $\alpha_X$ disappear in this limit.

\subsec{$Z$-configuration and its $\ap=0$ limit}

The result will be expressed in terms of the following kinematic variables:
\eqn\abcx{  \alpha_Z=-\,\langle 13\rangle[24][\,6|Z|5\rangle~,\qquad
\beta_Z=\langle 13\rangle[46][\,2|Z|5\rangle~,\qquad
\gamma_Z=\langle 51\rangle[24][\,6|Z|3\rangle~,}
where $\alpha_Z$, already defined in Eq. \kin, is listed for completeness. Then
\eqn\agz{\eqalign{
A& ( \,g^{-}_1,g^{+}_2, g^{-}_3,g^{+}_4,g^{-}_5,g^{+}_6)=A^Z=
{\rm Tr}(\,T^{a_1}T^{a_2}T^{a_3}T^{a_4}T^{a_5}T^{a_6})\
{(\sqrt{2}\,g_{Y\! M})^4\,\ap^5  \over s_5}\cr
&\ \times \lf(\,N^Z_1{\alpha_Z^2\over s_{13}^2s_{24}^2} ~+
N^Z_2{\beta_Z^2\over s_{13}^2} ~+N^Z_3{\gamma_Z^2\over s_{24}^2} ~+
N^Z_4{\alpha_Z\beta_Z\over s_{13}^2s_{24}} ~+
N^Z_5{\alpha_Z\gamma_Z\over s_{13}s_{24}^2} ~+
N^Z_6{\beta_Z\gamma_Z\over s_{13}s_{24}}\, \ri)\ ,}}
with the functions $N^Z$ written below:
\eqn\NZ{\eqalign{
N^Z_1&=-(s_1+s_2+s_3+s_6-t_2-t_3)\ (s_1+s_2+s_3+s_4-t_1-t_3)\ G^Z_1\cr
&-(s_1+s_2+s_3+s_6-t_2-t_3)\ [\ (s_5+s_6-t_2)\ G^Z_2-(s_4+s_5-t_1)\
G^Z_5-2\ s_5\ H_2^Z\ ]\cr
&+(s_1+s_2+s_3+s_4-t_1-t_3)\ [\ (s_5+s_6-t_2)\ G^Z_3-(s_4+s_5-t_1)\ G^Z_4-
2\ s_5\ H_3^Z\ ]\cr
&-(s_4+s_5-t_1)\ (s_5+s_6-t_2)\ G_6^Z-2\ s_5\ (s_5+s_6-t_2)\ H_1^Z-s_5\ L^Z_5\ ,\cr
N^Z_2&=-G^Z_5\ ,\cr
N^Z_3&=-G^Z_3\ ,\cr
N^Z_4&=-(s_1+s_2+s_3+s_4-t_1-t_3)\ (G_1^Z-G_4^Z+G_5^Z)
-(s_5+s_6-t_2)\ (G_2^Z -G_5^Z-G_6^Z)\cr
&+2\ (s_4-s_6-t_1+t_2)\ G_5^Z+2\ s_5\ H_2^Z\ ,\cr
N^Z_5&=-(s_1+s_2+s_3+s_6-t_2-t_3)\ (G^Z_1-G^Z_2+G_3^Z)-(s_4+s_5-t_1)\
(G_3^Z+G_4^Z-G_6^Z)\cr
&+2\ (s_5+s_6-t_2)\ G_3^Z+2\ s_5\ (H_1^Z-H_3^Z)\cr
N^Z_6&=-G^Z_1+G^Z_2+G^Z_4-G^Z_6\ .}}
The low-energy expansions of these functions are much more complicated in the
two previous cases:
\eqn\nz{\eqalign{
N^Z_1&=(s_1+s_2-t_1)\ (s_2+s_3-t_2)\
\lf(\fc{s_2}{s_1s_3t_3}+\fc{s_5}{s_4s_6t_3}+
\fc{t_1}{s_1s_4t_3}+\fc{t_2}{s_3s_6t_3}\ri)+\ldots\ ,\cr
N^Z_2&=(s_1+s_2-t_1)\ \lf(\fc{1}{s_1 s_3 t_3}+\fc{1}{s_1 s_4 t_3}+\fc{1}{s_1 s_4
   t_1}+\fc{1}{s_2 s_4 t_1}  \ri)+\ ,\cr
N^Z_3&=(s_2+s_3-t_2)\lf(\fc{1}{s_1 s_3 t_3}+\fc{1}{s_3 s_6 t_3}+\fc{1}{s_2 s_6
   t_2}+\fc{1}{s_3 s_6 t_2}  \ri)+\ldots\ ,\cr
N^Z_4&=(s_1+s_2-t_1)\lf(\fc{2 s_2}{s_1 s_3 t_3}+\fc{s_2}{s_1 s_4 t_3}+\fc{1}{s_1
   s_4}-\fc{s_5}{s_1 s_4 t_3}-\fc{s_6}{s_1 s_4
   t_3}+\fc{t_1}{s_1 s_4 t_3}-\fc{t_2}{s_1 s_3
   t_3}\ri.\cr
&\lf.+\fc{t_2}{s_3 s_6 t_3}+\fc{t_2}{s_4 s_6 t_3}+\fc{1}{s_1
   t_3}+\fc{s_3}{s_1 s_4 t_3}-\fc{1}{s_4 t_3}+\fc{t_2}{s_4 s_6
   s_2}   \ri)+\ldots\ ,\cr}}
$$\eqalign{N^Z_5&=(s_2+s_3-t_2)\lf(\fc{s_1}{s_3 s_6 t_3}+\fc{t_1}{s_2 s_4 s_6}+\fc{1}{s_3
   s_6}+\fc{t_1}{s_4 s_6 t_3}+\fc{t_2}{s_3 s_6 t_3}+\fc{1}{s_3
   t_3}-\fc{s_4}{s_3 s_6 t_3}\ri.\cr
&\lf.-\fc{s_5}{s_3 s_6
   t_3}+\fc{s_2}{s_3 s_6 t_3}-\fc{1}{s_6 t_3}-\fc{t_1}{s_3 t_3
   s_1}+\fc{t_1}{s_4 t_3 s_1}+\fc{2 s_2}{s_3
   t_3 s_1}    \ri)+\ldots\ ,\cr
N^Z_6&=\fc{2 s_2}{s_1 s_3 t_3}+\fc{s_2}{s_1 s_4 t_3}+\fc{s_2}{s_3
   s_6 t_3}+\fc{s_2}{s_4 s_6 t_3}+\fc{1}{s_1 s_4}+\fc{1}{s_3
   s_6}+\fc{2}{s_4 s_6}+\fc{s_5}{s_1 s_3 t_3}-\fc{s_6}{s_1 s_4
   t_3}\cr
&-\fc{t_1}{s_1 s_3 t_3}-\fc{t_2}{s_1s_3 t_3}-\fc{1}{s_4
   t_3}-\fc{s_4}{s_3 s_6 t_3}-\fc{1}{s_6 t_3}+\fc{t_3}{s_4 s_6
   s_2}-\fc{1}{s_4 s_2}-\fc{1}{s_6 s_2}+\ldots\ ,}$$
where dots represent terms suppressed by a factor of order
${\cal O}\big(\zeta(2)\ap^2\big)$ with respect to the leading (QCD)
contributions to the amplitude. The order $\zeta(2)\ap^2$
is presented in Appendix~B.

{}For this helicity configuration, the comparison of the $\ap=0$ limit with
QCD is a highly nontrivial and tedious exercise in spinor algebra which,
fortunately, has a happy end.
Most likely, another SUSY relation would be more efficient in handling this case.

More details on the functions $N_i$ are given in Appendix B.

\newsec{Summary and Outlook}

Together with the MHV amplitudes presented in Refs. \refs{\STi,\STii,\August},
the NMHV amplitudes presented in this work provide
the complete six--gluon disk amplitude.
As expected, the NMHV case is considerably more complex than MHV.
Six gluons are still manageable (as well as seven--gluon MHVs \August),
but clearly more efficient techniques need to be developed for handling
larger numbers of external gluons. To that end, some type of recursion
relations should be constructed, similar to Berends--Giele relations
\BerendsME\ in QCD and/or to the so called MHV or recursive
rules \refs{\CSW,\BrittoAP,\BrittoFQ}.
This is quite an involved task: all string excitations propagate in
intermediate channels of the disk diagram, therefore a part of the problem
is to extend string propagation off mass--shell.

In addition to possible phenomenological
applications of our results already stressed in the Introduction, we should
point out that the complete six--gluon string amplitude,
together with the previously obtained five-- and
four--gluon amplitudes (summarized in Refs.\refs{\STi,\STii}), provide all
information  necessary for constructing the non--Abelian Born--Infeld action
up to the order ${\cal O}(\ap^4F^6)$ in the gauge field strength $F$.
Furthermore, a direct comparison of type I disk amplitudes with two-loop
heterotic amplitudes -- a non-trivial test of type I-heterotic duality \BI\ --
becomes now possible.

It is interesting that multi-gluon disk amplitudes exhibit transcendentality
behavior in their low-energy $\ap$--expansions. Each power $\ap^n$ comes
with the factor $\zeta(n)$, a product of zeta functions or multiple zeta
values having transcendentality degree $n$.

\goodbreak
\vskip 5mm
\centerline{\noindent{\bf Acknowledgements} }

We are grateful to Lance Dixon for a very useful correspondence that allowed
improving the original version of this paper.
This work  is supported in part by the European Commission under
Project MRTN-CT-2004-005104.
The research of T.R.T.\ is supported in part by the U.S.
National Science Foundation Grant PHY-0600304.
St.St. would like to thank the Galileo Galilei Institute for Theoretical
Physics in Firenze and the Albert--Einstein--Institut in Potsdam for
hospitality and INFN and AEI for partial support during completion of this
work.
He is grateful to Hermann Nicolai for inviting him to the AEI.
T.R.T.\ thanks Max Planck Institute in Munich and
High Energy Theory Group at Harvard University for
their kind hospitality. He is deeply indebted to Dieter L\"ust for a timely
invitation to Munich.
Any opinions, findings, and conclusions or
recommendations expressed in this material are those of the authors and do
not necessarily reflect the views of the National Science Foundation.

\vfill
\goodbreak
\appendix\appA{Hypergeometric functions $L_i, G_j$ and their $\ap$ expansions}

In this Appendix, we collect the hypergeometric functions
$L_i$ and $G_i$ describing the auxiliary amplitudes
\ssssgg, \ssssss\ and \agg.

\subsec{Helicity configuration $Y$}
\noindent
The functions $L_i$ entering \ssssgg\
are:
\eqn\Lfunction{\eqalign{
L^Y_1&=s_{23}\ L^Y_4-
\INT \lf(s_{23}+\fc{1-s_{23}}{x}\ri) \fc{\Ic(x,y,z)}{xyz(1-y)(1-z)}\ ,\cr
L^Y_2&=-s_{13}\ L^Y_4-\INT \lf(1-s_{13}+\fc{s_{13}}{x}\ri)\
\fc{\Ic(x,y,z)}{(1-y)(1-z)}\ ,\cr
L^Y_3&=s_{12}\ L^Y_4+s_{12}\ \INT\ \fc{1-x}{1-z}\ \fc{\Ic(x,y,z)}{x(1-y)(1-xy)}\ ,\cr
L^Y_4&=\INT\ \fc{(1-x)\ \Ic(x,y,z)}{xz(1-y)(1-xy)}\ .}}
Their low-energy expansions are:
\eqn\expansions{\eqalign{
L^Y_1&=-\fc{1}{s_2 s_4}-\fc{1}{s_2 s_5}-\fc{1}{s_2
   s_6}+\fc{1}{s_4 s_6}+\fc{s_1}{s_2 s_4 t_1}+\fc{s_1}{s_2 s_5 t_1}+
\fc{s_3}{s_2 s_5 t_2}+\fc{s_3}{s_2 s_6t_2}+\fc{t_3}{s_2s_4 s_6}+\ldots\ ,\cr
L^Y_2&=-\fc{1}{s_2 s_4}-\fc{1}{s_2
   s_5}+\fc{1}{s_4 s_6}+\fc{s_1}{s_2 s_4 s_6}+\fc{s_1}{s_2 s_4 t_1}+\fc{s_1}{s_2
   s_5 t_1}-\fc{t_1}{s_2 s_4 s_6}+\ldots\ ,\cr
L^Y_3&=\fc{s_1}{s_2s_4s_6}+\fc{s_1}{s_2s_4t_1}+\fc{s_1}{s_2s_5t_1}+\ldots\ \ \
,\ \ \ L^Y_4=\fc{1}{s_2s_4s_6}+\ldots\ .}}
The function $L_5$ entering \ssssss\ is:
\eqn\Lfunctioni{\eqalign{
L^Y_5=-\INT&
\lf[\lf(1-s_{13}\fc{1-xyz}{z(1-xy)}\ri)\lf(1-s_{24}\fc{1-yz}{1-y}\ri)\ri.\cr
&+\fc{1}{x^2}\lf(1-s_{14}\fc{1-yz}{z(1-y)}\ri)\lf(1-s_{23}\fc{1-xyz}{1-xy}\ri)\cr
&\lf.+\fc{(1-yz)(1-xyz)}{xz(1-y)(1-xy)}
\lf(s_{12}s_{34}-s_{14}s_{23}-s_{13}s_{24}\ri)\ri]\ \fc{\Ic(x,y,z)}{y(1-z)^2}\ .}}
It has  the following $\ap$--expansion:
\eqn\lyexpp{\eqalign{
L^Y_5&=\fc{1}{s_2s_4}\ \lf(s_1-s_3+s_5-\fc{s_1s_5}{t_1}-t_1\ri)\
+\fc{1}{s_2s_6}\ \lf(-s_1+s_3+s_5-\fc{s_3s_5}{t_2}-t_2\ri)\cr
&+\fc{1}{s_4s_6}\ \lf(s_1+s_3-s_5-\fc{s_1s_3}{t_3}-t_3\ri)
-\fc{1}{s_2s_4s_6}(s_3t_1+s_1t_2+s_5t_3-t_1t_2-t_1t_3-t_2t_3)\cr
&-\fc{1}{s_2s_5} \lf(\fc{s_1s_4}{t_1}+
\fc{s_3s_6}{t_2}-t_3\ri)+
\fc{2}{s_2} \lf(1-\fc{s_1}{t_1}-\fc{s_3}{t_2}\ri)+\ldots\ .}}
The functions $G_i$ entering \agg\ are:
\eqn\gys{\eqalign{
G^Y_1&=\int\limits_0^1 dx\int\limits_0^1 dy\ \int\limits_0^1 dz\
\lf(\fc{1-s_{14}}{x}+s_{14}\ri)\ \fc{xy\ \Ic(x,y,z)}{(1-xy)(1-xyz)}\ ,\cr
G^Y_2&=-s_{34}\int\limits_0^1 dx\int\limits_0^1 dy\ \int\limits_0^1 dz\
\fc{\Ic(x,y,z)}{xyz}\ ,\cr
G^Y_3&=\int\limits_0^1 dx\int\limits_0^1 dy\ \int\limits_0^1 dz\
\lf(1-s_{24}+\fc{s_{24}}{x}\ri)\ \fc{\Ic(x,y,z)}{yz(1-xy)(1-xyz)}\ ,\cr
G^Y_4&=-s_{12}\int\limits_0^1 dx\int\limits_0^1 dy\ \int\limits_0^1 dz\
\fc{y (1-x)^2\ \Ic(x,y,z)}{x(1-y)(1-xy)(1-yz)(1-xyz)}\ ,\cr
G^Y_5&=\int\limits_0^1 dx\int\limits_0^1 dy\ \int\limits_0^1 dz\
\lf(1-s_{13}+\fc{s_{13}}{x}\ri)\ \fc{y\ \Ic(x,y,z)}{(1-y)(1-yz)}\ ,\cr
G^Y_6&=\int\limits_0^1 dx\int\limits_0^1 dy\ \int\limits_0^1 dz\
\lf(\fc{1-s_{23}}{x}+s_{23}\ri)\ \fc{\Ic(x,y,z)}{xyz(1-y)(1-yz)}\ .}}
Their low-energy expansions are
\eqn\gyexp{\eqalign{
G_1^Y&=\zeta(2)+\ldots\ \ \ ,\ \ \ G_2^Y=-\fc{s_3}{s_2s_6t_2}+\zeta(2)\
\lf(\fc{s_3s_4}{s_2s_6}+
\fc{s_3s_5}{s_2t_2}+\fc{s_3^2}{s_6t_2}\ri)+\ldots\ ,\cr
G_3^Y&=\fc{1}{s_2s_6}-\fc{s_3}{s_2s_6t_2}+\zeta(2)\
\lf(1-\fc{s_5}{s_2}-\fc{s_3}{s_6}
-\fc{t_3}{s_6}+\fc{s_3s_4}{s_2s_6}+\fc{s_3s_5}{s_2t_2}+
\fc{s_3^2}{s_6t_2}-\fc{s_4t_2}{s_2s_6}\ri)+\ldots\ ,\cr
G_4^Y&=-\fc{s_1}{s_2 s_4 t_1}+\zeta(2)\ \lf(\fc{s_1s_6}{s_2 s_4}+
\fc{s_1^2}{s_4 t_1}+\fc{s_1 s_5}{s_2t_1}\ri)+\ldots\ ,\cr
G_5^Y&=\fc{1}{s_2s_4}-\fc{s_1}{s_2s_4t_1}+\zeta(2)\
\lf(1-\fc{s_1}{s_4}-\fc{s_5}{s_2}
-\fc{t_3}{s_4}+\fc{s_1s_6}{s_2s_4}+\fc{s_1s_5}{s_2t_1}+\fc{s_1^2}{s_4t_1}-
\fc{s_6t_1}{s_2s_4}\ri)+\ldots\ ,\cr}}
$$\eqalign{
G_6^Y&=\fc{1}{s_2s_4}+\fc{1}{s_2 s_6}-\fc{s_1}{s_2 s_4 t_1}-
\fc{t_3}{s_2 s_4s_6}-\fc{s_3}{s_2s_6 t_2}+\zeta(2)
   \lf(1-\fc{s_1}{s_2}-\fc{s_3}{s_2}-\fc{s_5}{s_2}-\fc{s_1}{s_4}-\fc{t_3}{s_4}-
\fc{s_3}{s_6}\ri.\cr
&\hskip-0.75cm\lf.-\fc{t_3}{s_6}+\fc{s_1^2}{s_4 t_1}+\fc{s_1 s_6}{s_2s_4}
     +\fc{s_3s_4}{s_2 s_6}+\fc{s_3 s_5}{s_2t_2}
     +\fc{s_1 s_5}{s_2t_1}+\fc{t_3^2}{s_4 s_6}-\fc{s_6t_1}{s_2 s_4}
     -\fc{s_4 t_2}{s_2 s_6}+\fc{t_1 t_3}{s_2s_4}+\fc{t_2 t_3}{s_2s_6}
     +\fc{s_3^2}{s_6 t_2}\ri)+\ldots,}$$
where dots represent terms suppressed by a factor of order
${\cal O}\big(\zeta(3)\ap^3\big)$ with respect to the leading (QCD) contribution.

\subsec{Helicity configuration $X$}
\noindent
The functions $L_i$ entering \ssssgg\
are:
\eqn\LLfunction{\eqalign{
L^X_1&=-
\INT \lf(s_{23}+\fc{1-s_{23}}{x}\ri) \fc{\Ic(x,y,z)}{xy(1-z)(1-yz)}\ ,}}
$$\eqalign{
L^X_2&=-\INT \lf(1-s_{13}+\fc{s_{13}}{x}\ri)\ \fc{\Ic(x,y,z)}{(1-z)(1-yz)}\
,\cr
L^X_3&=s_{12}\ \INT\ \fc{1-x}{1-z}\ \fc{\Ic(x,y,z)}{x(1-yz)(1-xyz)}\ ,\cr
L^X_4&=\INT\ \fc{(1-x)\ \Ic(x,y,z)}{x(1-yz)(1-xyz)}\ .}$$
Their low-energy expansions are:
\eqn\expansionstil{\eqalign{
L^X_1&=-\fc{1}{s_2s_5}+\fc{s_1}{s_2s_5t_1}+\fc{s_3}{s_2s_5t_2}+\ldots\ \ \ ,\ \ \
L^X_2=-\fc{1}{s_2s_5}+\fc{s_1}{s_2s_5t_1}+\ldots\ ,\cr
L^X_3&=\fc{s_1}{s_2s_5t_1}+\ldots\ \ \ ,\ \ \ L^X_4=\ldots\ .}}
The function $L_5$ entering \ssssss\ is
\eqn\Lfunctioni{\eqalign{
L^X_5=-\INT&
\lf[\lf(1-s_{13}\fc{z(1-xy)}{(1-xyz)}\ri)\lf(1-s_{24}\fc{1-y}{1-yz}\ri)\ri.\cr
&+\fc{1}{x^2}\lf(1-s_{14}\fc{(1-y)z}{1-yz}\ri)\lf(1-s_{23}\fc{1-xy}{1-xyz}\ri)\cr
&\lf.+\fc{(1-y)(1-xy)z}{x(1-yz)(1-xyz)}
\lf(s_{12}s_{34}-s_{14}s_{23}-s_{13}s_{24}\ri)\ri]\ \fc{\Ic(x,y,z)}{y(1-z)^2}\ .}}
It has the following $\ap$--expansion:
\eqn\Expansionn{
L^X_5=\fc{t_3}{s_2s_5}-\fc{s_1s_4}{s_2s_5t_1}-\fc{s_3s_6}{s_2s_5t_2}+\ldots\ .}
where dots represent terms suppressed by a factor of order
${\cal O}\big(\zeta(2)\ap^2\big)$ with respect to the leading (QCD) contribution.

As already mentioned before $A^X_g=A^Y_g$, therefore the functions
$G_i^X=G_i^Y$, see Eq.\gys.

\subsec{Helicity configuration $Z$}
\noindent
The functions $L_i$ entering \ssssgg\
are:
\eqn\LLLfunction{\eqalign{
L_1^Z&=-\INT \lf(1-\fc{s_{23}}{1-x}\ri) \fc{\Ic(x,y,z)}{x^2yz(1-y)(1-z)}\ ,\cr
L_2^Z&=-s_{13}\INT \fc{\Ic(x,y,z)}{xz(1-x)(1-y)(1-z)}\ ,\cr
L_3^Z&=-\INT \lf(1-\fc{s_{12}}{xz}\ri)\
\fc{\Ic(x,y,z)}{(1-x)(1-y)(1-z)(1-xyz)}\ ,\cr
L_4^Z&=\INT\ \fc{\Ic(x,y,z)}{xz(1-x)(1-y)(1-xyz)}\ .}}
Their low-energy expansions are:
\eqn\expansionstill{\eqalign{
L_1^Z&=\fc{s_1}{s_2 s_4 t_1}+\fc{s_1}{s_2 s_5 t_1}+\fc{t_3}{s_2s_4 s_6}+
\fc{s_3}{s_2 s_5 t_2}+\fc{s_2}{s_3 s_5t_2}+\fc{s_3}{s_2 s_6 t_2}+\fc{s_2}{s_3
   s_6 t_2}+\fc{s_2}{s_3 s_6t_3}\cr
&+\fc{s_2}{s_4 s_6 t_3}+\fc{s_2}{s_1s_3 s_5}+\fc{s_2}{ s_1s_4 t_1}+\fc{s_2}{s_1s_5 t_1}
+\fc{s_2}{ s_1s_3 t_3}+\fc{s_2}{ s_1s_4 t_3}-\fc{1}{s_2 s_4}-\fc{1}{s_2 s_5}-
\fc{1}{s_2s_6}\cr
&+\fc{2}{s_4 s_6}+\fc{2}{s_4 t_1}+\fc{2}{s_5t_1}+\fc{2}{s_5 t_2}+\fc{2}{s_6 t_2}+
\ldots\ ,\cr
L_2^Z&=(s_1+s_2-t_1)\lf(\fc{1}{s_2 s_4 s_6}+\fc{1}{s_2 s_4 t_1}+\fc{1}{s_1s_5
t_1}+\fc{1}{s_2s_5 t_1}+\fc{1}{s_3 s_6 t_3}+\fc{1}{s_4 s_6t_3}+\fc{1}{ s_1s_3s_5}\ri.\cr
&\lf.\hskip2.5cm +\fc{1}{ s_1s_3 t_3}+\fc{1}{ s_1s_4
t_1}+\fc{1}{ s_1s_4 t_3}\ri)+\ldots\ ,\cr
L_3^Z&=\fc{s_1}{s_2 s_4 s_6}+\fc{s_1}{s_2 s_4 t_1}+\fc{s_1}{s_2
   s_5 t_1}+\fc{s_1}{s_3 s_6 t_3}+\fc{s_1}{s_4 s_6
   t_3}+\fc{t_2}{ s_1s_3 s_5}+\fc{s_2}{s_1s_4 t_1}+\fc{s_2}{ s_1s_5 t_1}+
\fc{s_6}{s_1s_3t_3}\cr
&+\fc{s_6}{ s_1s_4 t_3}-\fc{1}{s_1 s_5}+\fc{1}{s_3 s_5}-\fc{1}{s_1
   s_3}-\fc{1}{ s_1s_4}+\fc{2}{s_4 t_1}+\fc{2}{s_5 t_1}+\fc{2}{s_3
   t_3}+\fc{2}{s_4 t_3}+\ldots\ ,\cr
L_4^Z&=\fc{1}{s_2 s_4s_6}+\fc{1}{s_1 s_3 t_3}+\fc{1}{s_1 s_4 t_3}+\fc{1}{s_3 s_6 t_3}+
\fc{1}{s_4 s_6 t_3}+\ldots\ .}}
The function $L_5$ entering \ssssss\ is
\eqn\Lfunctioni{\eqalign{
L_5^Z=-\INT&
\lf[\fc{1}{(1-x)^2}\lf(1-\fc{s_{12}}{z}\ri)\lf(1-s_{34}\fc{(1-xy)(1-yz)}{(1-y)(1-xyz)}
\ri)\ri.\cr
&+\fc{1}{x^2}\lf(1-s_{14}\fc{1-yz}{(1-y)z}\ri)\lf(1-s_{23}\fc{1-xy}{1-xyz}\ri)\cr
&\lf.+\fc{(1-xy)(1-yz)}{xz(1-x)(1-y)(1-xyz)}
\lf(s_{12}s_{34}+s_{14}s_{23}-s_{13}s_{24}\ri)\ri]\ \fc{\Ic(x,y,z)}{y(1-z)^2}\ .}}
It has the following expansion:
\eqn\Expansionnn{\eqalign{
L_5^Z&=\fc{1}{s_2s_4}\ \lf(s_1-s_3+s_5-\fc{s_1s_5}{t_1}-t_1\ri)\
+\fc{1}{s_2s_6}\ \lf(-s_1+s_3+s_5-\fc{s_3s_5}{t_2}-t_2\ri)\cr
&+\fc{1}{s_4s_6}\ \lf(s_1+s_3-s_5-\fc{s_1s_3}{t_3}-t_3\ri)
-\fc{1}{s_2s_4s_6}(s_3t_1+s_1t_2+s_5t_3-t_1t_2-t_1t_3-t_2t_3)\cr
&+\fc{1}{s_3s_5}\ \lf(s_2-s_4+s_6-\fc{s_2s_6}{t_2}-t_2\ri)\
+\fc{1}{s_1s_3}\ \lf(-s_2+s_4+s_6-\fc{s_4s_6}{t_3}-t_3\ri)\cr
&+\fc{1}{s_1s_5}\ \lf(s_2+s_4-s_6-\fc{s_2s_4}{t_1}-t_1\ri)
-\fc{1}{s_1s_3s_5}(s_4t_2+s_2t_3+s_6t_1-t_2t_3-t_1t_2-t_1t_3)\cr
&-\fc{1}{s_2s_5} \lf(\fc{s_1s_4}{t_1}+\fc{s_3s_6}{t_2}-t_3\ri)
 -\fc{1}{s_3s_6} \lf(\fc{s_2s_5}{t_2}+\fc{s_1s_4}{t_3}-t_1\ri)
 -\fc{1}{s_1s_4} \lf(\fc{s_3s_6}{t_3}+\fc{s_2s_5}{t_1}-t_2\ri)\cr
&+\fc{2}{s_1} \lf(1-\fc{s_6}{t_3}-\fc{s_2}{t_1}\ri)
+\fc{2}{s_2} \lf(1-\fc{s_1}{t_1}-\fc{s_3}{t_2}\ri)
+\fc{2}{s_3} \lf(1-\fc{s_2}{t_2}-\fc{s_4}{t_3}\ri)\cr
&+\fc{2}{s_4} \lf(1-\fc{s_3}{t_3}-\fc{s_5}{t_1}\ri)
+\fc{2}{s_5} \lf(1-\fc{s_4}{t_1}-\fc{s_6}{t_2}\ri)
+\fc{2}{s_6} \lf(1-\fc{s_5}{t_2}-\fc{s_1}{t_3}\ri)-
4\lf(\fc{1}{t_1}+\fc{1}{t_2}+\fc{1}{t_3}\ri)+\ldots\ ,}}
where dots represent terms suppressed by a factor of order
${\cal O}\big(\zeta(2)\ap^2\big)$ with respect to the leading (QCD) contribution.
Actually, up to the last term, which is invariant
under cyclic permutations, the leading contribution of \Expansionnn\ is the
cyclicized version of the expansion
of $L^Y_5$ given in \lyexpp. Indeed $L_5^Z$ is invariant
under cyclic permutations to all orders in $\ap$.

The functions $G_i$ entering \agg\ are:
\eqn\gzs{\eqalign{
G_1^Z&=\int\limits_0^1 dx\int\limits_0^1 dy\ \int\limits_0^1 dz\
\lf(\fc{1-s_{14}}{x}-\fc{s_{14}}{1-x}\ri)\ \fc{xy\ \Ic(x,y,z)}{(1-xy)(1-xyz)}\ ,\cr
G_2^Z&=\int\limits_0^1 dx\int\limits_0^1 dy\ \int\limits_0^1 dz\
\lf(\fc{1-s_{34}}{1-x}-\fc{s_{34}}{x}\ri)\ \fc{\Ic(x,y,z)}{yz(1-x)}\ ,\cr
G_3^Z&=s_{24}\ \int\limits_0^1 dx\int\limits_0^1 dy\ \int\limits_0^1 dz\
\fc{\Ic(x,y,z)}{xyz(1-x)(1-xy)(1-xyz)}\ ,\cr}}
$$\eqalign{
G_4^Z&=\int\limits_0^1 dx\int\limits_0^1 dy\ \int\limits_0^1 dz\
\lf(1-\fc{s_{12}}{x}\ri)\
\fc{y\ \Ic(x,y,z)}{(1-y)(1-xy)(1-yz)(1-xyz)}\ ,\cr
G_5^Z&=s_{13}\ \int\limits_0^1 dx\int\limits_0^1 dy\ \int\limits_0^1 dz\
\fc{y\ \Ic(x,y,z)}{x(1-x)(1-y)(1-yz)}\ ,\cr
G_6^Z&=\int\limits_0^1 dx\int\limits_0^1 dy\ \int\limits_0^1 dz\
\lf(\fc{1-s_{23}}{x}-\fc{s_{23}}{1-x}\ri)\ \fc{\Ic(x,y,z)}{xyz(1-y)(1-yz)}\ .}$$
Their low-energy expansions are
\eqn\Assumeexp{\eqalign{
G_1^Z&=-\fc{s_2+s_5-t_1-t_2}{ s_1  s_3t_3}+\ldots\ ,\cr
G_2^Z&=\fc{1}{s_3s_6}-\fc{1}{s_6t_3}-\fc{2}{s_6t_2}-\fc{s_2}{s_3s_6t_2}-
\fc{s_3}{s_2s_6t_2}
-\fc{s_4}{s_3s_6t_3}+\ldots\ ,\cr
G_3^Z&=s_{24}\ \lf(\fc{1}{s_2s_6t_2}+\fc{1}{s_3s_6t_2}+\fc{1}{s_1s_3t_3}+
\fc{1}{s_3s_6t_3}\ri)+\ldots\ ,\cr
G_4^Z&=\fc{1}{s_1s_4}-\fc{1}{s_4t_3}-\fc{2}{s_4t_1}-\fc{s_1}{s_2s_4t_1}-
\fc{s_2}{s_1s_4t_1}-\fc{s_6}{s_1s_4t_3}+\ldots\ ,\cr
G_5^Z&=s_{13}\ \lf(\fc{1}{s_1s_4t_1}+\fc{1}{s_2s_4t_1}+\fc{1}{s_1s_3t_3}+
\fc{1}{s_1s_4t_3}\ri)+\ldots\ ,\cr
G_6^Z&=\fc{1}{s_2s_4}+\fc{1}{s_2s_6}-\fc{2}{s_4s_6}-\fc{2}{s_4t_1}-\fc{2}{s_6t_2}
-\fc{s_1}{s_2s_4t_1} -\fc{s_2}{s_1s_4t_1} -\fc{s_2}{s_3s_6t_2} -\fc{s_3}{s_2s_6t_2}\cr
&-\fc{s_2}{s_1s_3t_3}-\fc{s_2}{s_1s_4t_3} -\fc{s_2}{s_3s_6t_3}
-\fc{s_2}{s_4s_6t_3} -\fc{t_3}{s_2s_4t_6} +\ldots\ ,}}
where dots represent terms suppressed by a factor of order
${\cal O}\big(\zeta(2)\ap^2\big)$ with respect to the leading (QCD) contribution.

\appendix\appB{Basis representation of the functions $N_i$}

In Ref.\Dan\ it was
shown that the full six--gluon amplitude can be expressed in a basis of six
multiple hypergeometric functions. In \doubref\STi\STii\ we introduced
a specific basis, $\{F_1,\ldots,F_6\}$
to describe the MHV amplitude.
These six hypergeometric functions are represented by generalized Euler integrals:
\eqn\Functs{\eqalign{
&F_1=\INT\fc{\Ic(x,y,z)}{xyz}\qquad \hskip 1.3cm F_2=\INT\fc{\Ic(x,y,z)}{z\ (1-xy)}\cr
&F_3=\INT\fc{\Ic(x,y,z)}{1-xyz}\qquad \hskip 1.3cm
F_4=\INT\fc{y\ \Ic(x,y,z)}{(1-xy)(1-yz)}}}
$$\eqalign{
&F_5=\INT\fc{\Ic(x,y,z)}{(1-xy)(1-xyz)}\qquad
F_6=\INT\fc{\Ic(x,y,z)}{(1-yz)(1-xyz)}\ .}$$
In order to make contact between NMHV and MHV amplitudes, it would be
desirable to express the NMHV functions, in particular the functions $N_i$
that enter the final result, in
terms of this basis. Indeed, some NMHV functions are related to this basis in
a simple way. For example:
\eqn\simpleEXPP{\eqalign{
G^Y_1&=s_6\ F_2+(s_1-s_5-s_6+t_2)\ F_3-s_1\ F_5\ \ \ ,\ \ \ G^Y_2=-s_3\ F_1\ ,\cr
G^Y_3&=-(s_3-t_2)\ F_1+(s_6-t_3)\ F_2-(s_3+s_6-t_2-t_3)\ F_3-s_1\ F_5\ ,}}
however other relations are more complicated. Here, we focus on  the functions $N_i$,
see Eqs.\NY, \NX\ and \NZ, which determine the NMHV amplitudes \agy, \agx\
and \agz, respectively.
\subsec{Helicity configuration $Y$}
\eqn\NNY{\eqalign{
N_1^Y&=s_1s_3(s_4t_3)^{-1}\ \{\ s_2 s_5 t_2\  F_1+s_2 s_5 (s_4+s_5-t_1)\ N_0 -
[s_1 s_5+(s_4-t_1) (s_5+t_3)]\ \tilde N_0\cr
&-s_5[s_3 s_4 -s_2 (s_3-t_3)]\ F_2-s_5 s_1(s_2-t_1)\
F_3-s_1 s_4s_5\ F_5-s_3 s_4 t_3 \ (F_3-F_5) \cr
&+(s_5+t_3)\  [s_4t_2\ F_2+s_2s_5\ F_3+t_1\ (s_4 -t_1)F_3+ s_4 \ (s_4 -t_1)
(F_4-F_5)+s_4s_5\ F_4]   \ \}\ ,\cr
N^Y_2&=s_1(s_4t_1)^{-1}\ \{\ s_6 t_2\ F_1+[s_6 (s_5-t_1)-s_4 (s_5-t_2)]\ N_0 -
(s_1-s_3)\  \tilde N_0\cr
&+[s_5 s_6+t_1(s_1-t_3)]\ F_3 -(s_1-s_3) (s_4+s_6) \ F_3-s_4 s_5\  (F_3+F_4-F_5)\cr
&-(s_3-t_3)\ [t_2\ F_3+ (s_1-s_3)\ (F_3-F_5)+(s_5-t_1)\ (F_4-F_5)]\ \}\ ,\cr
N^Y_3&=s_3\ F_1\ ,\cr
N_4^Y&=s_1s_4^{-1}\ \{\ t_2^2\  F_1+t_2\ (s_4+s_5-t_1)\ N_0
-(s_1-s_3-s_5-t_3) \ \tilde N_0 +s_3 t_2 \ (F_2-F_3)\cr
&-t_2 t_3\  F_2+2 s_3 s_4 F_3+(s_1-s_3-s_5-t_3)\ (t_1-t_2) F_3+
s_4\  (s_1+s_3-s_5-t_3)\ (F_4-F_5)\ \},\cr
N_5^Y&=s_3s_4^{-1}\ \{\ t_1 t_2 F_1-
(s_3 s_4 -s_4s_5-s_4 t_3+t_1 t_3)\ (F_2-F_3)+ (s_1 s_4 +s_4 s_5+ s_3 t_1)\ F_2\cr
&+t_1\ (s_4+s_5-t_1)\ N_0+(s_1 s_4-s_1t_1+s_5t_1)\  F_3 \ \}\ ,\cr}}
$$\eqalign{
N_6^Y&=s_4^{-1}\ \{\ t_2\ t_3\ F_1+[s_4\ (s_1+s_3-s_5)+t_3\ (s_5-t_1)]\ N_0
+ s_3\ t_3\ F_2-t_3^2\ (F_2-F_3)\cr
&-(s_1-s_5)\ t_3\ F_3\ \}\ ,}$$
with the definitions:
\eqn\with{\eqalign{
N_0&= \fc{s_6}{s_2}\ (F_2-F_3)\cr
&-\fc{s_3-s_5+t_1-t_3}{s_2}\ (F_3+ F_4)-
\fc{s_1-s_3+s_5-t_1}{s_2}\ F_5+\fc{s_1+s_3-s_5-t_3}{s_2}\ F_6\ ,\cr
\tilde N_0&=s_6\ (F_2-F_3)+(s_1-s_3+t_2)\ F_3+(s_4+s_5-t_1)\ (F_4-F_5)-
(s_1-s_3)\ F_5\ .}}

\subsec{Helicity configuration $X$}

\eqn\NNX{\eqalign{
N^X_1&=-s_1 s_3\ \{\
s_6\ (F_2+F_3)+s_1\ (F_3-F_5)+ (s_5-t_1)\ (F_3+F_4-F_5)-s_5\ F_6\ \},\cr
N_2^X&=s_1(s_4t_1)^{-1}\ \{\ s_6\ t_2\ F_1+
[ s_6\ (s_5-t_1)-s_4\ (s_5-t_2)]\ N_0- (s_1 -s_3)\ \tilde N_0\cr
&-(s_4+s_6)\ (s_1-s_3)\ F_3-(s_3-t_3)(s_5-t_1)\ (F_4-F_5)+
(s_1-s_3)\ (s_3-t_3)\ (F_3-F_5)\cr
&-s_4 s_5\ (F_3+F_4-F_5)  +[s_5 s_6+t_1\ (s_1-t_3)- t_2\ (s_3-t_3)] \ F_3\ \}\ ,\cr
N^X_3&=s_3\ F_1\ ,\cr
N^X_4&=s_1\ s_4^{-1}\ \{\ s_4\ t_2\ N_0+t_3\ \tilde N_0-t_1\ t_3\ F_3-s_4\
t_3\ (F_4-F_5)+2\ s_3\ s_4\ (F_3+F_4-F_5)\ \}\ ,\cr
N^X_5&=s_3\ \{\ t_1\ N_0+t_3\ (F_2-F_3)+2\ s_1\ F_3\ \},\cr
N_6^X&=s_4^{-1}\ \{\ t_2\ t_3\ F_1+[s_4\ (s_1+s_3-s_5)+t_3\ (s_5-t_1)]\ N_0
+ s_3\ t_3\ F_2-t_3^2\ (F_2-F_3)\cr
&-(s_1-s_5)\ t_3\ F_3\ \}\ .}}

\subsec{Helicity configuration $Z$}

\eqn\NNZ{\eqalign{
s_1s_3s_6&(s_1+s_2-t_1)^{-1}(s_2+s_3-t_2)^{-1}\ N^Z_1=
[s_6 (s_1 s_3 s_5+s_2 s_4 s_6+s_3 s_6 t_1+s_1 s_4 t_2)] \tilde F_1\cr
&+\lf\{ -s_2 s_4 s_5 s_6+\left[-s_3 t_1^2+s_3 (s_1+s_2+s_4)
   t_1+s_4 \left(s_2^2+s_1 (t_2-s_3)\right)\right] s_6+s_1 s_4\ri.\cr
&\lf.\times(s_2+s_3-t_2) t_2 \ri\}\tilde F_2+\lf\{-s_2 s_6 s_5^2-
   s_5s_6 [s_2 (s_2+s_3+s_6-t_1-2 t_2)-s_1 s_3]\ri.\cr
&+(s_2+s_3) s_6^2 t_1-s_1 t_2(s_2+s_3-t_2)
   (s_1-s_3+s_4-t_1+t_2)-s_6 \left[(s_4-t_1) s_2^2\ri.\cr
&-(t_1 (s_3-2t_2)+s_4 t_2) s_2+s_3 t_1 (-s_3+s_4-t_1+t_2)+s_1
   \left(s_2^2+2 (s_3-t_2) s_2+s_3^2\ri.\cr
&\lf.\lf.\lf.+t_2^2+s_3 (-s_4+t_1-2 t_2)\right)\right]   \ri\}\ \tilde F_3\cr
&\hskip-1cm+\lf\{-(s_2+s_3-t_2) [s_2 s_5 s_6+(s_1 (s_2+s_3-t_2)-(s_2+s_3)
   t_1) s_6+s_1 t_2 (s_1-s_3-t_1+t_2)]    \ri\}\tilde F_4\cr
&+\lf\{s_2 s_6 s_5^2+[-(s_2+s_3) s_6 t_1+(s_1+s_2) s_6
   (s_3-t_2)+s_1 t_2 (-s_2-s_3+t_2)] s_5+s_6 t_1\ri.\cr
&\hskip-1cm\times \lf.((s_1+s_2+s_3) t_2-s_3 (s_2+s_3))+s_1 (s_2+s_3-t_2)
   (s_6 (s_2+s_3-t_2)+t_2 (s_1-s_3+t_2))   \ri\}\tilde F_5\cr
&+\lf\{ (s_2+s_3-t_2) [(s_2+s_3) s_6 (s_1-t_1)+s_1 (s_1-s_3-t_1)
   t_2+s_5 (s_2 s_6+s_1 t_2)]
   \ri\}\tilde F_6\ ,\cr\cr
s_1s_2s_3t_1&(s_1+s_2-t_1)^{-1}\ N^Z_2=
\lf\{s_2 (s_3+s_4) s_6 t_1+(s_1+s_2) s_3 s_6 t_3\ri\}\ \tilde F_1\cr
&+\lf\{-s_2 s_3 t_1^2+(s_1+s_2) s_3 (-s_1 s_2+s_3 s_4-s_4 s_5+s_1 s_6+s_2
t_3)\ri.\cr
&\lf.+t_1 (-s_1 s_3 s_6+s_2 (2 s_1 s_3+2 s_3 s_4+s_2 (s_3+s_4)-s_4 s_5-s_3
t_3))\ri\}\ \tilde F_2\cr
&+\lf\{s_2 t_1^2 (s_2+2 s_3+s_5+s_6-2 t_2)+s_3 [-
(-s_1+s_2+s_3-s_4+s_5+s_6-2 t_2)\ri.\cr
&\times(-s_3+s_5)-(s_2+s_5-t_2) t_3](s_1+s_2) +t_1 [-s_2 s_5^2-s_2 s_5 (s_2+s_6-2 \
t_2)\ ,\cr}}
$$\eqalign{
&\lf.+s_1 (-s_2^2+s_3 (s_3+s_6-t_2)+s_2 (-2 s_3+t_2))+s_2 (2 s_3^2+s_2
(s_3-s_4)\ri.\cr
&\lf.+2 s_3 s_6+s_4 t_2+s_3 (-2 s_4-3 t_2+t_3))]\ri\}\ \tilde F_3\cr
&+\lf\{(-s_1+s_3-s_5+t_1) [s_1 s_3+s_2 (s_3+t_1)] (s_2+s_3-t_2)\ri\}\ \tilde F_4\cr
&+\lf\{-s_2 t_1^2 (s_3+s_5-t_2)+(s_1+s_2) s_3 (-s_3+s_5)
(-s_1+s_3+s_5-t_2)\ri.\cr
&\lf.+s_2 t_1 [s_5^2+s_1 (s_2+3 s_3-t_2)+2 s_3 (-s_3+t_2)-s_5 (s_3+t_2)]\ri\}\
\tilde F_5\cr
&+\lf\{(s_1-s_3+s_5-t_1) [s_1 s_3+s_2 (s_3+t_1)] (s_2+s_3-t_2)\ri\}\ \tilde F_6\ ,\cr\cr
s_1s_3t_3&(s_2+s_3-t_2)^{-1}\ N^Z_3=\lf\{s_2 (s_1+s_6) t_2+s_1 (s_2+s_3) t_3\ri\}\ F_1\cr
&+\lf\{-(s_1+s_6) (-s_2 s_3+s_3 s_4+s_1 s_6-s_5 s_6-s_4 t_2)\ri.\cr
&\lf.+[-s_5 s_6+s_1 (2 s_3+s_4+2 s_6-t_2)] t_3-s_1 t_3^2\ri\}\ F_2\cr
&+\lf\{(s_1+s_6) [-(s_1-s_5) (s_1+s_2-s_3+s_4+s_5-s_6-2 t_1)-(s_1+s_4-t_1) t_2]\ri.\cr
&+[s_1^2-s_5 (s_4+s_5-s_6-2 t_1)+s_3 t_1-s_2(s_3+s_5-t_2)+(s_4+s_5-2 t_1) t_2\cr
&\lf.+s_1 (2 s_2-2 s_3+s_4+s_5-2 s_6-3 t_1+2 t_2)] t_3+s_1 t_3^2\ri\}\ F_3\cr
&+\lf\{-(s_4+s_5-t_1) [(s_1+s_3-s_5) (s_1+s_6)+(-2 s_1+s_5-t_2) t_3]\ri\}\ F_4\cr
&+\lf\{(s_1-s_5) (s_1+s_6) (s_1-s_3+s_5-t_1)-[2 s_1^2-s_2 s_3+s_3
t_1\ri.\cr
&\lf.-(s_5-t_1) (s_5-t_2)+s_1 (-3 s_3+s_5-2 t_1+t_2)] t_3\ri\}\ F_5\cr
&+(s_4+s_5-t_1) [(s_1+s_3-s_5) (s_1+s_6)+(-2 s_1+s_5-t_2) t_3]\ F_6\ ,\cr\cr
s_1s_2s_3s_6&(s_1+s_2-t_1)^{-1}\ N_4^Z=
\lf\{-s_2 s_3 s_5 s_6-s_2 s_3 s_6^2+s_1 t_2 [s_2 (s_3+s_4)+s_3
t_3]\ri.\cr
&\lf.+s_2s_6 [-s_1 s_3+s_3^2+s_2 (s_3+2 s_4)-s_4 t_2+s_3 (s_4+t_1+t_3)]
\ri\}\ s_6 \tilde F_1\cr
&+\lf\{-s_2 s_3 s_6^2 (s_1+s_2-t_1)+s_1 (s_2+s_3) s_4 (s_2+s_3-t_2) t_2\ri.\cr
&-s_2 s_5 s_6 [s_1 s_3+2 \
s_3 s_4+s_2 (s_3+2 s_4)-s_3 t_1-s_4 t_2]+s_6 [s_1^2 s_3 (-s_2+t_2)\cr
&+s_2 [s_2^2 (s_3+2s_4)+s_2 (s_3^2-s_4 t_2+s_3 (3 s_4+t_3))+s_3 (s_3 (2
s_4-t_1)+s_4 (t_1-t_2)\cr
&\lf.-t_1 (t_1+t_3))]+s_1 [s_3 (s_4-t_1) t_2+
s_2 (s_3^2+s_4 t_2+s_3 (-s_4+2 t_1+t_2+t_3))]]\ri\}\ \tilde F_2\cr
&+\lf\{s_2 s_6^2 [s_1 s_3+s_2 (3 s_3+2 t_1)+(2 s_3+t_1) (s_3-t_2)]+s_2 s_5^2 s_6 (-2 \
s_2-s_3+t_2)\ri.\cr
&\lf.-s_1 (s_2+s_3) (s_2+s_3-t_2) t_2 (s_1-s_3+s_4-t_1+t_2)-s_2 s_5 s_6 [2 \
s_2^2-2 s_1 s_3+s_3^2\ri.\cr
&+s_2 (2 s_3-2 t_1-5 t_2)+s_6 (2 s_2+s_3-t_2)+t_2 (t_1+2 t_2)+s_3 \
(-s_4-3 t_2+t_3)]\cr
&+s_6 [s_1[s_1 s_3 (s_2-t_2)-2 s_2^3+4 s_2^2 (-s_3+t_2)+s_3 t_2 \
(s_3-s_4+t_1-t_2+t_3)\cr
&-s_2 [3 s_3^2+2 t_2^2+s_3 (-s_4+2 t_1-4 t_2+t_3)]]
+s_2 [2 s_3^3+s_2^2 (s_3-2 s_4+2 t_1)\cr
&+s_3^2 (-2 s_4+3 t_1-4 t_2)
-(s_4-2 t_1) t_2^2+s_2 [3 \ s_3^2+(3 s_4-5 t_1) t_2\cr
&\lf.-s_3 (3 s_4-4 t_1+4 t_2+t_3)]+s_3 [t_1^2+t_1 (-s_4-4 \
t_2+t_3)+t_2 (2 (s_4+t_2)+t_3)]]]   \ri\}\ \tilde F_3}$$
$$\eqalign{
&+\lf\{(s_2+s_3-t_2) [s_1 (s_2+s_3) (-s_1+s_3+t_1-t_2) t_2+s_2 s_5 s_6
(-2 (s_2+s_3)+t_2)\ri.\cr
&\lf.+s_6 [-2 s_1 s_2 (s_2+s_3)+s_2 (s_3+t_1) (2 (s_2+s_3)-t_2)+s_1 (2 s_2+s_3) t_2]]
\ri\}\ \tilde F_4\cr
&+\lf\{s_2 s_5^2 s_6 (2 (s_2+s_3)-t_2)+s_1 (s_2+s_3) (s_2+s_3-t_2) t_2
(s_1-s_3+t_2)\ri.\cr
&-s_5 [s_1 (s_2+s_3) (s_2+s_3-t_2) t_2+s_6 (s_1 (s_2+s_3) t_2+s_2 [2 (s_2+s_3)-t_2] \
(t_1+t_2))]\cr
&+s_6 [-s_2 (s_3+t_1) (s_3-t_2) (2 (s_2+s_3)-t_2)+s_1^2 s_3 t_2+s_1 [2 \
s_2^3+s_2^2 (6 s_3-4 t_2)\cr
&\lf.+s_3 t_2 (-s_3+t_2)+s_2 [4 s_3^2-5 s_3 t_2+t_2 (t_1+2 \
t_2)]]]   \ri\}\ \tilde F_5\cr
&+\lf\{ (s_1-s_3+s_5-t_1) (s_2+s_3-t_2) [s_2 s_6 (2 (s_2+s_3)-t_2)+s_1
(s_2+s_3) t_2]  \ri\}\ \tilde F_6\ ,}$$
$$\eqalign{
s_1s_2s_3s_6&(s_2+s_3-t_2)^{-1}\ N^Z_5=
\lf\{ s_6 [-s_1 s_2 s_4 s_5+s_2 s_6 (s_4 (s_1+2 s_2-t_1)+s_3 t_1)\ri.\cr
&\lf.+s_1 s_3 t_1 (s_2+t_3)+s_1 s_2 \
s_4 (s_1+s_2-s_3-s_4+t_2+t_3)]    \ri\}\ \tilde F_1\cr
&+\lf\{ -s_2 s_4 s_5 [2 (s_1+s_2) s_6-s_6 t_1+s_1 (s_2+s_3-t_2)]\ri.\cr
&+s_6 [(s_1+s_2) [(s_1+s_2) s_3+(-s_2+s_3) s_4] t_1-(s_1+s_2) s_3 t_1^2\cr
&+s_2 s_4 (2 s_1^2+2 s_2^2+s_1 (3 s_2-s_3+t_2))]\cr
&\lf.+s_1 s_4 (s_2+s_3-t_2) [s_3 t_1+s_2 (s_1+s_2-s_3-s_4+t_2+t_3)] \ri\}\ \tilde F_2\cr
&+\lf\{s_2 s_6^2 (s_1+2 s_2+s_3-t_1) t_1+s_2 s_5^2 s_6 (-s_1-2 s_2+t_1)\ri.\cr
&-s_1 (s_2+s_3-t_2) (s_1-s_3+s_4-t_1+t_2) [s_3 t_1+s_2 (s_1+s_2-s_3-s_4+t_2+t_3)]\cr
&-s_2 s_5 [s_6^2 (s_1+2s_2-t_1)-s_1 (s_2+s_3-t_2) (s_1-s_3+s_4-t_1+t_2)\cr
&+s_6 [-s_1^2+(2 s_2-t_1) (s_2+s_3-t_1-2 t_2)+s_1 (s_2+s_3-s_4-3 t_2+t_3)]]\cr
&+s_6 [s_1^2 (-(s_2+s_3) t_1+s_2 (s_2+s_3-t_2))-s_2 [2 s_2^2 (s_4-t_1)+t_1
[-s_3^2+(s_4-2 t_1) t_2\cr
&+s_3 (s_4+t_2)]+s_2 (-s_4 (t_1+2t_2)+t_1 (-2 s_3+t_1+4 t_2))]+
s_1 [-s_2^2 (s_3+s_4-t_2)\cr
&+s_3 t_1 \
(s_3-s_4+t_1-t_2+t_3)+s_2 [-s_3^2-2 t_1 t_2-t_2^2+s_4 (-t_1+t_2)+s_3 (s_4+2
t_2)\cr
&\lf.+t_1 (t_1+t_3)]]]     \ri\}\ \tilde F_3\cr
&+\lf\{(s_2+s_3-t_2) [s_2 s_5 [s_6 [-2 (s_1+s_2)+t_1]+s_1
(s_1-s_3-t_1+t_2)]\ri.\cr
&+s_6 [s_1^2 s_2+s_2 (2 s_2+s_3-t_1) t_1+s_1 (s_3 t_1+s_2
(-s_3+t_1+t_2))]\cr
&\lf.-s_1 (s_1-s_3-t_1+t_2) (s_3 t_1+s_2 (s_1+s_2-s_3-s_4+t_2+t_3))]\ri\}\ \tilde F_4\cr
&+\lf\{ (-s_1+s_3+s_5-t_2) [s_2 s_5 [2 (s_1+s_2) s_6-s_6 t_1+s_1
(s_2+s_3-t_2)]\ri.\cr
&+s_6 [-(s_1+s_2) (2 s_2+s_3) t_1+s_2 t_1^2+s_1 s_2 (s_2+s_3-t_2)]\cr
&\lf.-s_1 (s_2+s_3-t_2) (s_3 t_1+s_2 (s_1+s_2-s_3-s_4+t_2+t_3))]    \ri\}\ \tilde F_5\cr
&+\lf\{ -(s_2+s_3-t_2) [s_1 s_2 s_5^2-s_2 s_6 (s_1-t_1) (s_1+2
s_2+s_3-t_1)\ri.\cr
&-s_5 [s_2 s_6 (s_1+2 s_2-t_1)+s_1 (s_2+s_3) t_1+s_1 s_2
(s_2-s_4+t_2+t_3)]\cr
&\lf.-s_1 (s_1-s_3-t_1) (s_3 t_1+s_2 \
(s_1+s_2-s_3-s_4+t_2+t_3))]    \ri\}\ \tilde F_6\cr\ ,}$$
$$\eqalign{
s_1s_2s_3s_6&\ N^Z_6=s_6\
\lf\{-s_2 s_3 s_6^2+s_1 (s_2-s_4+t_3) (s_2 (s_3+s_4)+s_3 t_3)+s_6 [-s_1 s_3 \
(s_2+t_3)\ri.\cr
&\lf.+s_2 (s_2 (s_3+2 s_4)+s_4 s_5-s_4 (t_1+t_2)+s_3 t_3)]   \ri\}\ \tilde F_1\cr
&+\lf\{ -(s_1+s_2) s_3 s_6^2 (s_1+s_2-t_1)+s_1 (s_2+s_3) s_4 (s_2+s_3-t_2)
(s_2-s_4+t_3)\ri.\cr
&+s_6
[s_1^2 (s_2 (s_3+s_4)+s_3 t_3)+s_2 [s_2^2 (s_3+2 s_4)-s_4 (-s_3+s_5)
(s_3+s_5-t_2)\cr
&-s_2 (s_4 s_5+s_4 (t_1+t_2)+s_3 (-2 s_4+t_1-t_3))+t_1 (s_4 s_5-s_3 t_3)]+s_1 [2 s_2^2 \
(s_3+s_4)\cr
&\lf.-s_2 (s_3+s_4) t_1+2 s_2 s_3 t_3+s_3 (s_4 s_5-t_1 t_3)]]   \ri\}\ \tilde F_2\cr
&+\lf\{ -s_2 s_5^3 s_6+s_5^2 s_6 [s_1 s_3-s_2 (3 s_2+s_3+s_6-2 t_1-3 t_2)]\ri.\cr
&+s_6^2 [-s_1 s_3 (-s_2+s_3+t_1)+s_2 (s_3^2+s_2 (3 s_3+2 t_1)-t_1 (t_1+t_2)
-s_3 (t_1+2 t_2))]\cr
&-s_1 (s_2+s_3) (s_2+s_3-t_2) (s_1-s_3+s_4-t_1+t_2) (s_2-s_4+t_3)\cr
&+s_5 s_6 [s_1^2 s_2+s_1 (s_3 \
s_6+s_3 (s_3-t_1-2 t_2)+s_2 (2 s_3+s_4-t_1+t_2-t_3))\cr
&+s_2 [-2 s_2^2+s_3^2-s_2 (s_3+s_4-4 t_1-5 t_2)+(s_4-2 t_2) t_2
+s_6 (-2 s_2+s_3+2 t_1+t_2)\cr
&-t_1 (t_1+5 t_2)-s_3 \
(-2 t_1+t_3)]]-s_6 [s_1^3 s_3+s_1^2 [-s_2^2+s_2 (t_1+t_2)+s_3
(s_4-t_1+t_2-t_3)]\cr
&+s_1 [s_3^3+s_2 (s_2 (t_1-t_2)+t_2^2+t_1 (s_4-t_1-t_3))+s_3^2 (s_4+t_1-t_2-2
t_3)\cr
&+s_3 (s_2(s_4+t_1+t_2-2 t_3)+t_2 (-2 t_1+t_3))]+s_2 (-s_3^3-s_2^2 (s_3-2
s_4+2 t_1)\cr
&+(s_4-2 t_1) t_2 (t_1+t_2)+s_3^2 (s_4+3 t_2)+s_3 [t_1^2+t_1 (t_2-t_3)-
t_2 (s_4+2 t_2+t_3)]\cr
&\lf.+s_2 (-3 s_3^2-s_4 (t_1+3 t_2)+t_1 (t_1+5 t_2)+s_3 (2 s_4-3 t_1+4
t_2+t_3)))]\ri\}\ \tilde F_3\cr
&+\lf\{ (s_2+s_3-t_2) [-s_2 s_5^2 s_6+s_5 s_6 (s_1 (-s_2+s_3)+s_2 (-2 s_2+2
t_1+t_2))\ri.\cr
&+s_6 [s_1^2 (s_2+2 s_3)+s_2 (s_3+t_1) (2 s_2+s_3-t_1-t_2)+s_1 (-2 s_3 t_1+s_2 (2 \
s_3+t_2))]\cr
&\lf.+s_1 (s_2+s_3) (-s_1+s_3+t_1-t_2) (s_2-s_4+t_3)]    \ri\}\ \tilde F_4\cr
&+\lf\{ (-s_1+s_3+s_5-t_2) [s_2 s_5^2 s_6+s_5 s_6 (s_1 (s_2-s_3)+s_2 (2 s_2-2
t_1-t_2))\ri.\cr
&-s_6 (s_1^2 s_3+s_2 (s_3+t_1) (2 s_2+s_3-t_1-t_2)+s_1 (-s_2^2-s_3 t_1+s_2 \
(s_3+t_1+t_2)))\cr
&\lf.-s_1 (s_2+s_3) (s_2+s_3-t_2) (s_2-s_4+t_3)]    \ri\}\ \tilde F_5\cr
&\lf\{ (s_1-s_3+s_5-t_1) (s_2+s_3-t_2) [s_2 s_5 s_6+s_6 [-s_1 s_3+s_2 (2
s_2+s_3-t_1-t_2)]\ri.\cr
&\lf.+s_1 (s_2+s_3) (s_2-s_4+t_3)]    \ri\}\ \tilde F_6\ ,}$$
with:
\eqn\Fs{\eqalign{
\tilde F_3&=F_3+F_4-F_6\ ,\ \tilde F_4=F_6\ ,\ \tilde F_5=F_4\ ,\ \tilde
F_6=F_5\ ,\cr
\tilde F_2&=\fc{s_6}{s_4}\
F_2+\left(1+\fc{s_1}{s_4}-\fc{s_6}{s_4}-\fc{t_1}{s_4}+
\fc{t_2}{s_4 }-\fc{s_3}{s_4}\right)\
F_3+\left(1+\fc{s_5}{s_4}-\fc{t_1}{s_4}\right)\ F_4\cr
&+\left(-\fc{s_1}{s_4}-\fc{s_5}{s_4}+\fc{t_1}{s_4}+\fc{s_3}{s_4}\right)\
F_5-F_6\ ,}}
$$\eqalign{
\tilde F_1&=\fc{s_2 t_2}{s_4 t_3}\ F_1+\left(-\fc{s_2}{s_4}+\fc{s_3 s_2}{s_4
   t_3}-\fc{s_3}{t_3}+\fc{s_5 s_6}{s_4 t_3}-\fc{s_1 s_6}{s_4
   t_3}+\fc{t_2}{t_3}\right) F_2\cr
&+\left(-\fc{s_1^2}{s_4 t_3}+\fc{s_6
   s_1}{s_4 t_3}+\fc{2 t_1 s_1}{s_4 t_3}-\fc{t_2 s_1}{s_4
   t_3}-\fc{s_2 s_1}{s_4 t_3}+\fc{s_3 s_1}{s_4
   t_3}-\fc{s_1}{t_3}+\fc{s_5}{s_4}-\fc{t_1}{s_4}+\fc{s_2}{s_4}+
\fc{s_5^2}{s_4 t_3}\ri.\cr
&\lf.+\fc{s_2 s_5}{s_4 t_3}-\fc{s_3 s_5}{s_4t_3}+
\fc{s_5}{t_3}-\fc{s_5 s_6}{s_4 t_3}-\fc{2 s_5 t_1}{s_4
   t_3}+\fc{t_1 t_2}{s_4 t_3}-\fc{t_2}{t_3}\right)\ F_3\cr
&+\left(\fc{s_5^2}{s_4 t_3}+\fc{s_5}{s_4}-\fc{t_1 s_5}{s_4
   t_3}-\fc{s_1 s_5}{s_4 t_3}-\fc{s_3 s_5}{s_4
   t_3}+\fc{s_5}{t_3}-\fc{t_1}{s_4}-\fc{s_1}{t_3}-\fc{s_3}{t_3}+
  \fc{s_1 t_1}{s_4 t_3}+\fc{s_3 t_1}{s_4 t_3}\right)\ F_4\cr
&+\left(\fc{s_1^2}{s_4 t_3}-\fc{t_1 s_1}{s_4 t_3}-\fc{s_3
   s_1}{s_4 t_3}-\fc{s_5^2}{s_4 t_3}+\fc{s_3 s_5}{s_4
   t_3}+\fc{s_5 t_1}{s_4 t_3}\right)\ F_5\cr
&+\left(-\fc{s_5^2}{s_4t_3}-\fc{s_5}{s_4}+\fc{t_1 s_5}{s_4 t_3}+\fc{s_1 s_5}{s_4
   t_3}+\fc{s_3 s_5}{s_4
   t_3}-\fc{s_5}{t_3}+\fc{t_1}{s_4}+\fc{s_1}{t_3}+\fc{s_3}{t_3}-
\fc{s_1 t_1}{s_4 t_3}-\fc{s_3 t_1}{s_4 t_3}\right)\ F_6\ .}$$
Finally, we list the next order of the expansions \nz.
It comes with a factor of $\zeta(2)$:
  \eqn\expZ{\eqalign{
N_1^Z&=\ldots+\zeta(2)\ (s_1 + s_2 - t_1) (s_2 + s_3 - t_2)\ \lf\{-\fc{t_1^2}{s_1 s_4}+
\fc{t_1}{s_1}+\fc{t_1}{s_4}-\fc{s_3 t_1}{s_1 t_3}-\fc{s_6t_1}{s_4 t_3}-
\fc{t_2^2}{s_3 s_6}\ri.\cr
&\lf.-\fc{s_2 s_5}{s_1
   s_3}+\fc{t_2}{s_3}+\fc{t_2}{s_6}-\fc{s_2 s_5}{s_4 s_6}-\fc{s_2 s_4}{s_1
   t_3}-\fc{s_1 s_5}{s_4 t_3}-\fc{s_2 s_6}{s_3 t_3}-\fc{s_1 t_2}{s_3
   t_3}-\fc{s_4 t_2}{s_6 t_3}-\fc{s_3
   s_5}{s_6 t_3}-1\ri\}+\ldots\ ,\cr
N_2^Z&=\ldots+\zeta(2)\ (s_1 + s_2 - t_1)\ \lf\{-\fc{s_1}{s_4 t_1}-\fc{s_6}{s_2 s_4}+
\fc{2}{s_4}-\fc{s_5}{s_2 t_1}-\fc{s_2}{s_4t_1}-\fc{s_6}{s_3 t_3}-\fc{s_6}{s_4
t_3}-\fc{s_5}{s_3 s_1}\ri.\cr
&\lf.-\fc{t_1}{s_4
   s_1}-\fc{t_3}{s_4 s_1}-\fc{s_5}{t_1 s_1}-\fc{s_3}{t_3 s_1}-\fc{s_4}{t_3
   s_1}+\fc{2}{s_1}\ri\}+\ldots\ ,\cr
N_3^Z&=\ldots+\zeta(2)\ (s_2 + s_3 - t_2)\ \lf\{-\fc{s_1}{s_3 t_3}-\fc{t_2}{s_3 s_6}-
\fc{t_3}{s_3 s_6}+\fc{2}{s_3}-\fc{s_4}{s_2s_6}+\fc{2}{s_6}-\fc{s_5}{s_2 t_2}-
\fc{s_5}{s_3 t_2}\ri.\cr
&\lf.-\fc{s_2}{s_6t_2}-\fc{s_3}{s_6 t_2}-\fc{s_6}{s_3 t_3}-\fc{s_4}{s_6 t_3}-\fc{s_5}{s_3
   s_1}-\fc{s_4}{t_3 s_1}\ri\}+\ldots\ ,\cr
N_4^Z&=\ldots+\zeta(2)\ (s_1 + s_2 - t_1)\lf\{-\fc{s_3^2}{s_1 t_3}
-\fc{t_1 s_3}{s_1 s_4}+\fc{s_3}{s_1}+\fc{s_3}{s_4}-\fc{s_2
   s_3}{s_1 t_3}-\fc{s_4 s_3}{s_1 t_3}+\fc{s_5 s_3}{s_1 t_3}+\fc{s_6
   s_3}{s_1 t_3}-\fc{s_6 s_3}{s_4 t_3}\ri.\cr
&-\fc{t_1 s_3}{s_1 t_3}-\fc{t_2
   s_3}{s_6 t_3}+\fc{s_3}{t_3}-\fc{t_1^2}{s_1 s_4}-\fc{t_2^2}{s_2 s_6}+\fc{2
   s_2}{s_1}-\fc{2 s_5}{s_1}-\fc{s_5}{s_4}-\fc{2 s_6}{s_4}+\fc{s_5 t_1}{s_1
   s_4}+\fc{s_6 t_1}{s_1 s_4}+\fc{t_1}{s_1}-\fc{s_2 t_1}{s_1 s_4}\cr
&+\fc{2
   t_1}{s_4}-\fc{t_1 t_2}{s_2
   s_4}-\fc{t_2}{s_1}+\fc{t_2}{s_2}+\fc{t_2}{s_4}-\fc{s_2 t_2}{s_4 s_6}+\fc{2
   t_2}{s_6}-\fc{t_1 t_3}{s_1 s_4}-\fc{t_2 t_3}{s_4
   s_6}+\fc{t_3}{s_4}-\fc{s_1}{s_4}+\fc{s_2}{s_4}+\fc{s_6^2}{s_4 t_3}\cr
&-\fc{2 s_2
   s_4}{s_1 t_3}+\fc{s_5 s_6}{s_4 t_3}+\fc{s_1 s_6}{s_4 t_3}-\fc{s_2
   s_6}{s_4 t_3}-\fc{s_6}{t_3}-\fc{s_6 t_1}{s_4 t_3}+\fc{s_4 t_2}{s_1
   t_3}-\fc{s_1 t_2}{s_4 t_3}-\fc{s_4 t_2}{s_6 t_3}-1-\fc{t_2^2}{s_6 s_3}-\fc{2
   s_2 s_5}{s_1 s_3}\cr
&\lf.+\fc{s_5 t_2}{s_1 s_3}+\fc{t_2}{s_3}-\fc{2 s_2 s_6}{t_3
   s_3}-\fc{s_1 t_2}{t_3 s_3}+\fc{s_6
   t_2}{t_3 s_3}\ri\}+\ldots\ ,\cr}}
$$\eqalign{N_5^Z&=\ldots+\zeta(2)\ (s_2 + s_3 - t_2)\ \lf\{-\fc{s_1^2}{s_3
t_3}-\fc{t_2 s_1}{s_3
s_6}+\fc{s_1}{s_3}+\fc{s_1}{s_6}+\fc{s_4
   s_1}{s_3 t_3}+\fc{s_5 s_1}{s_3 t_3}-\fc{s_6 s_1}{s_3 t_3}-\fc{t_1
   s_1}{s_4 t_3}\ri.\cr
&-\fc{t_2 s_1}{s_3 t_3}-\fc{s_2 s_1}{s_3 t_3}-\fc{s_4
   s_1}{s_6 t_3}+\fc{s_1}{t_3}-\fc{t_1^2}{s_2 s_4}-\fc{t_2^2}{s_3 s_6}-\fc{2
   s_5}{s_3}+\fc{t_1}{s_2}-\fc{t_1}{s_3}+\fc{2 t_1}{s_4}-\fc{s_2 t_1}{s_4
   s_6}+\fc{t_1}{s_6}-\fc{t_1 t_2}{s_2 s_6}\cr
&+\fc{t_2}{s_3}+\fc{s_4 t_2}{s_3
   s_6}+\fc{s_5 t_2}{s_3 s_6}-\fc{s_2 t_2}{s_3 s_6}+\fc{2 t_2}{s_6}-\fc{t_1
   t_3}{s_4 s_6}-\fc{t_2 t_3}{s_3 s_6}+\fc{t_3}{s_6}+\fc{2
   s_2}{s_3}+\fc{s_2}{s_6}-\fc{s_3}{s_6}-\fc{2
   s_4}{s_6}-\fc{s_5}{s_6}-\fc{s_4}{t_3}\cr
&-\fc{2 s_2 s_6}{s_3 t_3}+\fc{s_6
   t_1}{s_3 t_3}-\fc{s_6 t_1}{s_4 t_3}-\fc{s_3 t_1}{s_6 t_3}-\fc{s_4
   t_2}{s_6 t_3}+\fc{s_4^2}{s_6 t_3}-\fc{s_2 s_4}{s_6 t_3}+\fc{s_3 s_4}{s_6
   t_3}+\fc{s_4 s_5}{s_6 t_3}-1-\fc{t_1^2}{s_4 s_1}-\fc{2 s_2 s_5}{s_3
   s_1}\cr
&\lf.+\fc{s_5 t_1}{s_3 s_1}+\fc{t_1}{s_1}-\fc{2 s_2 s_4}{t_3 s_1}-\fc{s_3
   t_1}{t_3 s_1}+\fc{s_4 t_1}{t_3
   s_1}\ri\}+\ldots\ ,\cr
N_6^Z&=\ldots+\zeta(2)\ \lf\{-\fc{s_2^2}{s_4 s_6}-\fc{2 s_5 s_2}{s_1 s_3}-
\fc{t_1 s_2}{s_1 s_4}-\fc{t_2
   s_2}{s_3 s_6}-\fc{2 t_3 s_2}{s_4 s_6}+\fc{2 s_2}{s_1}+\fc{2 s_2}{s_3}+\fc{3
   s_2}{s_4}+\fc{3 s_2}{s_6}-\fc{s_3 s_2}{s_1 t_3}\ri.\cr
&-\fc{2 s_4 s_2}{s_1 t_3}-\fc{2
   s_6 s_2}{s_3 t_3}-\fc{s_6 s_2}{s_4 t_3}-\fc{s_1 s_2}{s_3 t_3}-\fc{s_1
   s_2}{s_4 t_3}-\fc{s_3 s_2}{s_6 t_3}-\fc{s_4 s_2}{s_6 t_3}-\fc{s_5^2}{s_1
   s_3}-\fc{t_3^2}{s_4 s_6}-\fc{3 s_6}{s_4}+\fc{s_5 t_1}{s_1 s_3}\cr
&+\fc{s_6
   t_1}{s_1 s_4}-\fc{t_1}{s_3}-\fc{t_1}{s_4}+\fc{s_5 t_2}{s_1
   s_3}-\fc{t_2}{s_1}+\fc{s_4 t_2}{s_3 s_6}-\fc{t_2}{s_6}-\fc{t_1 t_3}{s_1
   s_4}-\fc{t_2 t_3}{s_3 s_6}+\fc{3 t_3}{s_4}+\fc{3
   t_3}{s_6}-\fc{s_1}{s_4}-\fc{s_3}{s_6}\cr
&-\fc{3 s_4}{s_6}+\fc{s_6^2}{s_4
   t_3}+\fc{s_1}{t_3}+\fc{s_3}{t_3}+\fc{s_1 s_4}{s_3 t_3}-\fc{s_4 s_5}{s_1
   t_3}+\fc{s_3 s_6}{s_1 t_3}-\fc{s_5 s_6}{s_3 t_3}+\fc{s_1 s_6}{s_4
   t_3}+\fc{s_4 t_1}{s_1 t_3}+\fc{s_6 t_1}{s_3 t_3}+\fc{s_4 t_2}{s_1
   t_3}\cr
&\lf.+\fc{s_6 t_2}{s_3 t_3}+\fc{s_4^2}{s_6 t_3}+\fc{s_3 s_4}{s_6
   t_3}+\fc{s_1}{s_2}+\fc{s_3}{s_2}+\fc{s_5}{s_2}+\fc{s_6 t_1}{s_4 s_2}+\fc{s_4
   t_2}{s_6 s_2}-\fc{t_1 t_3}{s_4
   s_2}-\fc{t_2 t_3}{s_6 s_2}\ri\}+\ldots\ .}$$

\listrefs
\end